# Emissions and cost tradeoffs of time-matched clean electricity procurement under inter-annual weather variability – case study of hydrogen production

Michael Giovanniello[1], Dharik S. Mallapragada[2]*

1. MIT Energy Initiative, Massachusetts Institute of Technology, Cambridge, MA 02139

2. Chemical and Biomolecular Engineering Department, Tandon School of Engineering, New York University, Brooklyn, NY 11201

* Corresponding author; Email: dharik.mallapragada@nyu.edu

## Abstract

Regulators and voluntary corporate sustainability efforts are increasingly adopting time-matching requirements (TMRs) for clean electricity procurement for large loads, such as data centers, and electricity-intensive fuel production, such as hydrogen. We use a stochastic capacity expansion model (CEM) framework to assess how inter-annual weather variability affects the cost, composition, and emissions of procurement-driven infrastructure to meet annual and hourly TMRs using the case study of a grid-connected hydrogen producer in Texas. Our approach, which relies on co-optimizing investments and hourly operations over nine weather scenarios, reveals that hourly TMR comes at a higher cost premium compared to annual TMR than previously estimated by single-scenario deterministic modeling, while emissions outcomes remain directionally consistent. Demand flexibility and partial hourly TMR (80–90%) lower the cost premium while preserving emissions benefits. We further examine how binding renewable portfolio standards (RPS) interact with TMR costs and emissions outcomes. When an RPS is applied to non-$H_2$ electricity demand, annual TMR reduces emissions comparably to hourly TMR at a lower cost. Incorporating $H_2$-related electricity demand directly into the RPS constraint, rather than imposing a separate TMR, achieves similar emissions outcomes at still lower cost, suggesting that TMR-based clean electricity procurement—particularly hourly matching—offers limited additional value in regions with stringent grid decarbonization policies.

## 1. Introduction

Electricity consumers—both large and small—are showing increasing interest in managing their electricity-related carbon footprint (also known as Scope 2 emissions) through additional procurement of clean electricity. By the end of 2024, voluntary procurement by consumers in the U.S. amounted to over 70 GW of variable renewable energy (VRE) generation capacity[1] – for context, total installed solar and wind capacity by end of 2023 was 95 GW and 149 GW, respectively.[2] As electricity demand for data centers supporting artificial intelligence (AI) applications is expected to grow rapidly,[3] it could also accelerate corporate clean energy procurement as firms look to meet their sustainability goals.

In addition to voluntary procurements, regulatory requirements are also driving electricity consumers to procure clean electricity. For example, electricity-based hydrogen ($H_2$) producers in the U.S. and E.U. must meet specific eligibility criteria to receive emissions-indexed production tax credits.[4,5] These include matching their electricity consumption with an equivalent amount of clean electricity from newly added, or so-called "additional," generation capacity—initially on an annual basis, and eventually on an hourly basis. Variants of these so-called time-matching requirements (TMRs) are also being considered in ongoing revisions of electricity-related emissions accounting protocols that many private sector firms adopt to manage and report their electricity-related emissions footprint[6].

Previous assessments of alternative clean electricity procurement strategies have used two methodological approaches: (a) *Demand-centric modeling,*[7–11] which evaluates the clean electricity investments—often co-located with demand—to meet consumer demand, and potentially offset grid electricity consumption or its associated emissions, while treating the consumer as a price-taker; and (b) *Grid-centric modeling,*[5,12–16] which examines the power system impacts of consumer-driven clean electricity procurement by co-optimizing grid-wide investments and operations to meet both aggregate demand and consumer clean electricity targets under various grid scenarios. Both modeling approaches have been extensively used to study the economics and emissions impacts of electricity-based $H_2$ production.[17–21] The discussion below, however, focuses on studies that consider clean electricity procurement strategies—irrespective of the end-use consumer—of which $H_2$ production is a common but not exclusive application. We also exclude discussion of systems involving co-located VRE generation and electrolytic $H_2$ production without grid connectivity, as the alternative clean electricity procurement strategies considered here are not relevant for those configurations.

To date, demand-centric modeling studies have focused on the costs of more stringent hourly time-matching vs. less stringent annual time-matching, as well as the impact of technological and spatial diversification.[7–9] For example, Casa Ferrus et al.[9] compare the levelized cost of hydrogen (LCOH) production under two procurement strategies—sourcing VRE supply from a single asset or from a spatially diversified portfolio of assets—and find cost savings with the portfolio approach. While many such studies rely on a single weather year to model VRE generation and system operations, a few demand-centric analyses have also accounted for VRE resources' inter-annual variability.[7,8] Such inter-annual variation has been shown to substantially influence investment needs—particularly for energy storage—in studies focused on power system decarbonization using 100% VRE supply.[22–24] Inter-annual VRE variation is particularly important for clean electricity procurement since consumers often are looking to sign multi-year contracts with VRE generators.

Many grid-centric modeling studies use capacity expansion models (CEMs) to quantify the power system-wide impacts and consumer-level costs and emissions under alternative clean electricity procurement strategies. Unlike demand-centric models, grid-centric analyses using CEMs capture competition among consumers for limited clean electricity resources (e.g., VRE) and evaluate how this competition affects clean electricity procurement outcomes, as well as system-level cost and emissions outcomes. These models also account for multiple revenue streams available to grid-connected VRE resources, including those arising from contracts with individual consumers and from sales of energy and capacity to the grid. As a result, they can provide a more realistic quantification of *additionality*—the causal relationship between consumer procurement and clean electricity deployment. Additionality measures whether clean electricity procurement actions actually cause new clean electricity generation to be added to the power system, beyond what would have been built anyway. For instance, Giovanniello et al.[12] used a CEM based framework to highlight how alternate definitions of additionality can change the emissions outcomes associated with electricity-based $H_2$ production that employs annual time-matching. The study found that weaker additionality definition, which classify any new generation as additional, can result in higher emissions. In contrast, more stringent definitions, that require that the new generation should not have been built without the consumer-driven TMR constraint, were found to result in emissions outcomes comparable to those achieved with hourly TMR[12]. Here, it is important to note that definitions of additionality vary across studies, depending on whether they are tracking emissions, generation or revenue impacts of clean electricity procurement and how they are computed. See Giovanniello et al.[12] for a discussion on the latter aspect of defining additionality.

Grid-centric modeling studies have identified several insights ito TMR as a strategy for clean electricity procurement: a) hourly matching is likely to lead to lower system-wide emissions than annual matching but is more expensive[12–14,25]; b) grid-level policies, such as binding requirements on shares of low-carbon or VRE generation, lower the emissions impact of the less granular annual matching without significantly impacting its cost[12,16,25]; c) the availability of clean firm resources and long-duration energy storage reduces the cost premium of more granular hourly matching[13,25,26], potentially paving the way for their scale-up through early adoption by consumers using such procurement strategies[27]; and d) the region-specific impacts of spatial and temporal matching requirements can vary significantly due to prevailing grid conditions and constraints on short-term renewable expansion[5,16].

The studies cited above use models that cosider only a single weather scenario for annual power system operations and do not account for inter-annual variations in VRE availability. In practice, however, consumers are interested in clean electricity procurement over longer time frames (e.g. 10 or 20 years), for which inter-annual variation in weather and its impact on VRE availability becomes relevant. Sizing clean electricity procurements based on a single weather scenario could lead to excess generation relative to demand in years of higher VRE availability and deficits in years with lower than anticipated VRE availability. These studies may also undervalue storage resources and diversification of the VRE resource portfolio, which help to to manage temporal mismatches between demand and the output of any single clean electricity resource. As noted earlier, while inter-annual weather variation has been examined in the context of islanded $H_2$ production system design[7,8] and broader grid decarbonization studies,[22–24] its effect on the system-wide impacts of clean electricity procurement strategies has, to our knowledge, not been previously considered.

Here, we investigate the impact of inter-annual VRE variation on the investments needed to meet either annual or hourly TMR for a grid-connected electricity-based $H_2$ producer to meet baseload $H_2$ demand. We use a CEM to evaluate the cost-optimal grid and procurement-driven infrastructure investments in VRE capacity, battery energy storage, $H_2$ storage, and electrolyzer capacity that are needed to meet demand and enforced TMRs (annual or hourly), as well as system operations. We consider both a) a single weather scenario ("Deterministic") and b) nine alternative weather scenarios ("Stochastic"). We then assess the robustness of the optimized infrastructure mix by testing its performance against 10 weather scenarios that are not included in the investment planning phase. Our case study, based on the Texas grid, explores a range of technological and policy scenarios involving different levels of energy storage availability and varying degrees of compliance with the enforced TMR for electrolytic hydrogen demand. We also examine how grid-wide decarbonization policies—specifically renewable portfolio standards (RPS) applied to existing demand—affect system emissions outcomes and the composition of clean electricity procurement portfolios for meeting annual and hourly TMRs under inter-annual VRE variability. We also evaluate the impact of incorporating additional electricity demand from $H_2$ production into an existing RPS policy rather than via separate TMR-based procurement. Finally, we undertake a revenue analysis of VRE resources deployed under scenarios with and without RPS to demonstrate how revenue metrics can be used to assess the additionality of resources in real-world settings.

# 2. Methods

## 2.1 Model overview

Our analysis uses the open-source CEM, DOLPHYN,[28] which is formulated as linear program (LP) for this study, to minimize the investment and operation cost of electricity and $H_2$ infrastructure while respecting constraints related to a) system operation including meeting specified hourly $H_2$ and electricity demands, b) technology operation, and c) policy requirements either at the technology or system level (See Table S1). DOLPHYN is capable of modeling transmission constraints, however, as a simplification, we ignore electricity and hydrogen transmission constraints in this study. This effectively assumes that demand and supply are balanced at the system level without accounting for any costs/constraints associated with transmission of electricity and $H_2$ between supply and demand. We provide a full description of the key model constraints in Table S1 in the supporting information (SI), and the model is available Github,[28,29] along with solution approach (S1.5).

Our analysis relies on two different CEM versions: a) *a deterministic model*, where we co-optimize investments and operations while considering system operations based on a single weather scenario at an hourly resolution and b) *a stochastic model*, where we co-optimize investments and operations while considering system operations based on nine different weather scenarios at an hourly resolution. Throughout the analysis "weather scenario" refers to the weather year that corresponds to availability time series for wind and solar resources, and exogenous electricity and $H_2$ demand are held constant across weather scenarios.

Eq. 1 shows the objective function of the stochastic model, which is composed of two parts: (1) the annualized investment cost of new resources ($k \in K$, first term) and (2) the expected operating costs for each weather scenario ($s \in S$ ), which includes annual fixed (2nd term) and variable operating costs (3rd term) for both existing and new resources as well as costs for load shedding (4th term). In Eq. 1, $K$ is set of all new resources, $T$ set of hours of the year, $G$ is the set of all existing and new resources, and $NSE$ is the cost associated with non-served energy for demand sources $D$, which are grid load and $H_2$ demand. Note that the model objective function in the model covers both electricity and hydrogen sectors, however for simplicity we have written the objective function in a sector agnostic manner. For the stochastic model, the annual operating costs of each weather scenario are weighted equally in the model objective function, i.e. $\sigma_s = 1/N_{scen}$, where $N_{scen}$ is the number of weather scenarios considered. For the deterministic model, $\sigma_s = 1$.

$$\text{Objective} = \sum_{k \in K} cap_k * inv_cost_k + \sum_{s \in S} \sigma_s \sum_{t \in T} \left( \sum_{g \in G} (fixed_cost_g(s,t) + var_cost_g(s,t)) + NSE(s,t)_{d \in D} \right) \quad (1)$$

Eqs. 2-3 describe the constraints enforcing hourly and annual time-matching of electricity consumption for $H_2$ production. For each hour, Eq. 2 requires that the electrolyzer power consumption, equal to tonnes of $H_2$ produced in that hour times the power required per tonne ($\lambda^{Ely}$) and the degree of hourly compliance ($\alpha_{TMR}$) must be less than or equal to generation from contracted VRE generation (i.e. PPA resources, $g \in TMR_g$) + net injection from set of eligible battery storage ($b \in TMR_b$ ). Here, $\alpha_{TMR}$ reflects the minimum fraction of $H_2$-related electricity consumption that needs to be matched by supply from PPA resources, and is a measure of degree of compliance to the TMR constraint.

$$\sum_{g \in TMR_g} gen_{g,t} + \sum_{s \in TMR_b} dischg_{s,t} - chg_{s,t} \geq \alpha_{TMR}\, gen_t^{ELY} \beta^{ELY} \; \forall t \in T \quad (2)$$

The annual TMR constraint, modeled by Eq. 3, enforces that the sum of annual generation from eligible set of VRE resources ($g \in TMR_g$ ) must be equal to annual electrolyzer electricity consumption.

$$\sum_{g \in TMR_g} \sum_{t \in T} gen_{g,t} + \sum_{s \in TMR_b} \sum_{t \in T} (dischg_{s,t} - chg_{s,t}) = \sum_{t \in T} gen_t^{ELY} \beta^{ELY} \quad (3)$$

Per Eq. 2, it is possible that the generation in some hours exceeds the electrolyzer power consumption to be matched, making the constraint non-binding. Effectively, in these hours, contracted VRE resources are generating electricity in excess of the contractual requirement. Following the approach of Zeyen et al.,[16] we included a constraint to limit the quantity of these "excess" electricity sales from contracted VRE resources (i.e. PPA resources) to the grid under the hourly time-matching requirement. Such a constraint is meant to discourage resources that would have been built for the grid to be designated as PPA resources in the optimal solution and thereby reduce model degeneracy. Practically, this constraint introduces a stronger operational relationship between PPA resources and electrolyzer by ensuring that majority of electricity generated by the VRE resources is contracted by the electrolyzer. In this way, it can be interpreted as a way to increase the degree of additionality of the PPA resources by minimizing the revenues obtained from electricity sales. Eq. 4 restricts the quantity of electricity sales from contracted VREs to the grid at 120% (β=0.2) of annual electrolyzer demand, which translates into a 20% excess sales allowance. The assumed value of β is based on assumptions in prior CEM studies evaluating time-matching requirement based electricity procurement.[16,25]

$$\sum_{t \in T} \left( \sum_{g \in TMR_g} gen_{g,t} + \sum_{s \in TMR_b} dischg_{s,t} - chg_{s,t} \right) \leq (1 + \beta) \sum_{t \in T} gen_t^{ELY} \beta^{ELY} \quad (4)$$

## 2.2 Case study description

We use a case study derived from generator and load conditions managed by the Electric Reliability Council of Texas (ERCOT) in 2021, where the existing generation capacity is dominated by natural gas (52.2 GW), followed by onshore wind (35.1 GW), solar (9.1 GW), coal (7 GW), and nuclear (5.0 GW) (see Table S6). Exogeneous electricity demand data assumptions are the same what was used in our previous publication (see SI Figure 1-2 of Giovanniello et al.[12]), with peak demand of 75.7 GW, and annual generation of 388.9 TWh, reflecting 2021 electricity demand patterns for the ERCOT grid.

Our choice of $H_2$ production as the case study is motivated by two factors. First, as noted earlier, clean electricity procurement requirements analogous to TMR have been explicitly contemplated in policies supporting electrolytic $H_2$ production in both the US and EU, making it a directly policy-relevant application. Second, electrolytic $H_2$ production offers the potential for flexible electricity consumption when deployed alongside $H_2$ storage, making it a particularly instructive case study for examining how demand flexibility shapes the cost and emissions outcomes of alternative procurement strategies.

To represent a consumer interested in clean electricity procurement, we model 1 GW of hourly $H_2$ demand (18.4 tonnes of $H_2$ per hour) to be met solely via electrolyzer-based $H_2$ production paired with $H_2$ storage. The modeled constant hourly $H_2$ demand is intended to reflect consumption patterns of existing industrial $H_2$ consumers, many of which are located in Texas. For context, the modeled $H_2$ demand is representative of a medium to large-scale ammonia production facility (440 $tH_2$/day, which is equivalent to >2500 tonne $NH_3$/day). Flexible electrolyzer operation is allowed in all scenarios (see S1.1), which previous analysis confirmed is the most economic operating mode[8,12,16] and is likely to be the preferred approach in practice. Note that we do not explicitly model the end-use $H_2$ consumer. Instead, our analysis is agnostic to the end-use industrial process consuming the produced $H_2$, which by definition of the baseload $H_2$ supply requirement, is not affected by electrolyzer or electric grid operation patterns. Electrolyzer cost and performance assumptions are summarized in Table S3 and reflect those of a PEM electrolyzer[30], which have greater operational flexibility than commercially available alkaline electrolyzers.

## 2.3 Technology assumptions

Investment and operating costs assumptions for new generation resources are from the 2022 NREL Annual Technology Baseline[31,32] (Table S1). Table S2 summarizes investment and operating costs assumptions for electrolyzer and gaseous $H_2$ storage – the same assumptions used in our previous study[12]. Model runs with retirement and expansion of grid resource capacity allow specifically for retirement of existing natural gas and coal and new construction of wind, solar, battery storage as well as new natural gas generation capacity if needed. Nuclear, hydro, and biomass are not subject to retirement or new construction. The model includes two Li-ion battery sizing decision variables — power capacity and energy capacity — linked by a constraint that bounds the energy-to-power ratio between 0.15 and 12 hours.

## 2.4 VRE availability data inputs

Hourly solar and wind availability profiles for different weather scenarios are constructed using empirical and simulated data from the ERCOT Hourly Wind and Solar Generation Profiles dataset.[33] The dataset includes empirical generation for solar and wind plants from 1980 to 2021. For plants built after 1980, historical generation profiles back to 1980 are simulated based on spatially granular historical weather data. The dataset also includes simulated generation profiles for plants that were under construction in 2020.

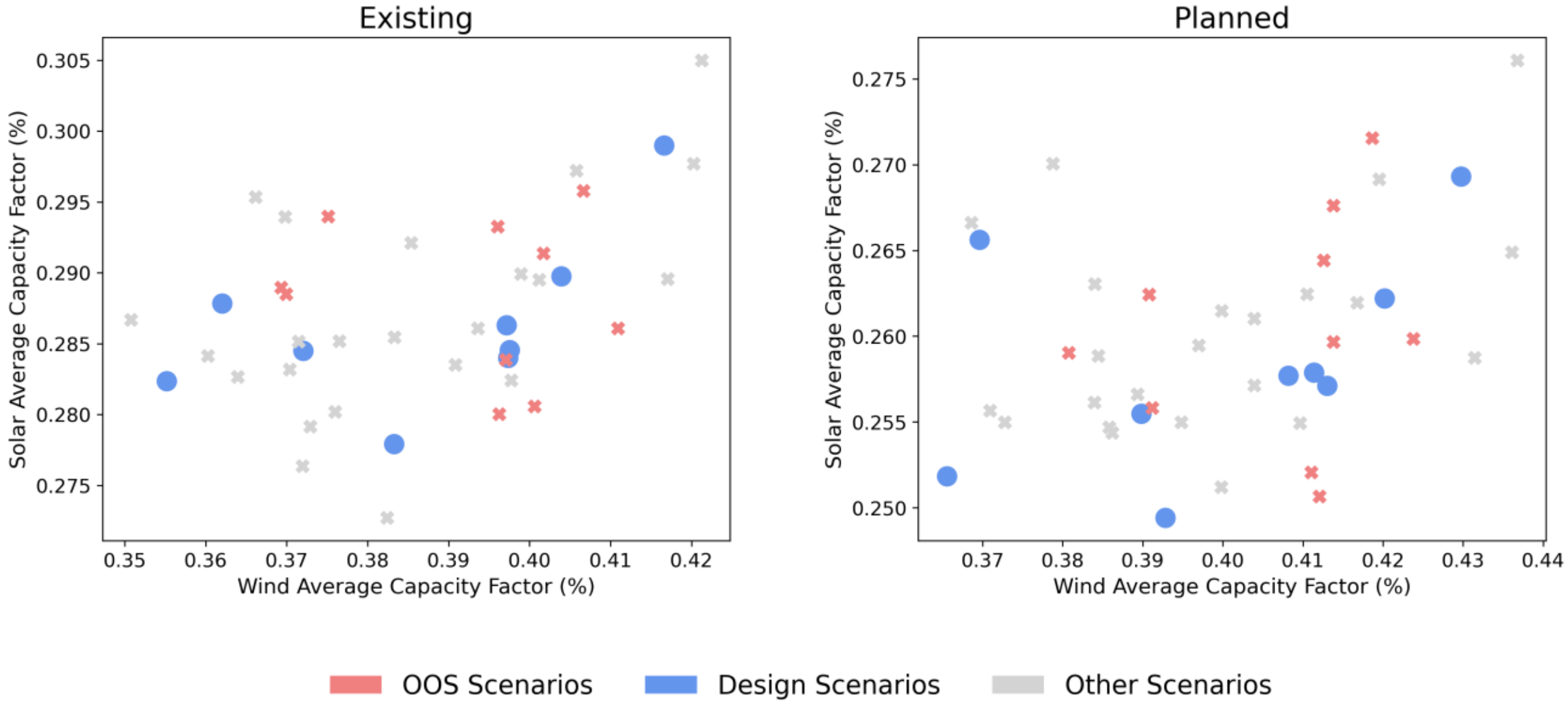

*Figure 1. Scatter plot of annual averaged capacity factors of wind (x-axis) and solar (x-axis) resources in ERCOT (1980-2021).[33] The nine VRE profiles used for the in-sample analysis are shown as large blue circle; profiles for the out-of-sample analysis are red 'X'; all remaining samples, which were used in the clustering analysis, are shown as gray 'X's. Figure S9 visualizes the same data organized by VRE resource type.*

We divided wind and solar resources into two groups– existing resources and new resources that could be expanded. For both resources, we created spatially aggregated capacity-weighted resource availability profiles by summing generation and dividing by total capacity for both existing and planned resources for each year. The result is four availability time series per weather scenario – one for existing and one for planned for both solar and wind – where the existing and planned profiles are used for existing and new build resources in the model, respectively (see S3.1 for details). Figure 1 highlights the inter-annual variation in annual average capacity factor for the four VRE resource types considered in the analysis, while Figures S10-S11 highlight the variation in hourly capacity factor data for each resource type.

Solving the stochastic model with 41 weather scenarios is computationally challenging with commercial optimization solvers like Gurobi and the available computing resources (see S1.5) when also considering a granular representation of the existing generation fleet (40 existing generators). To address this, we applied k-means clustering to identify nine representative weather scenarios (blue markers in Figure 1; see Section S3.1 for details), where each representative scenario is defined as the historical year closest to the centroid of its respective cluster. These nine weather scenarios are used in the in-sample analysis, which includes the nine deterministic model runs that each consider one weather scenario and a stochastic model run that considers all nine scenarios. Out-of-sample weather scenarios were selected by randomly sampling 10 of the remaining weather scenarios, which are highlighted by the red markers in Figure 1.

### 2.5 Scenarios evaluated and metrics of interest

We assess the effects of annual and hourly TMRs under various scenarios of model setup, weather scenario, technology availability, and policy design, summarized in Table 1. We quantify our findings in terms of impacts on model investments (power, electrolyzer and $H_2$ storage capacity), emissions, and cost.

Emissions impacts are quantified as the difference in grid-level emissions associated with system-wide power generation in the case of $H_2$ demand minus the emissions in the identical scenario without $H_2$ demand, divided by total annual $H_2$ production. This metric can be thought of as the consequential emissions impacts of $H_2$ production with clean electricity procurement using TMRs. It should be noted that this metric does not account for the avoided emissions from displacing any existing fossil-based hydrogen production. Since these avoided emissions are independent of the procurement strategy employed, they do not affect comparisons across scenarios but should be considered when interpreting consequential grid emissions estimates in absolute terms. In addition, our emissions metric only includes $CO_2$ emissions from fuel combustion associated with coal and NG-based electricity generation and does not account for upstream emissions related to fuel production, manufacturing, or non-$CO_2$ greenhouse gas emissions. Incorporating these elements would affect the emissions impacts of the various scenarios evaluated here (Table 1), though directional trends are likely to remain consistent given the predominance of fuel combustion emissions.

*Table 1. Description of model scenarios evaluated and their corresponding labels used in later figures.*

| Description of evaluated scenarios | Scenario labels |
|---|---|
| Deterministic (D) model with annual (A) or hourly TMR constraint with 100% compliance for different weather years (1-9) | Di-A for i = 1, …, 9<br>Di-H for i = 1, …, 9 |
| Stochastic (S) model with annual (A) or hourly TMR with 100% compliance | S-A, S-H |
| Stochastic (S) model with hourly TMR constraint serving atleast a fraction of demand (y) and limits on $H_2$ storage availability, equal to x hours of rated $H_2$ demand | S-H-y-xL Or S-H-y for y= 80%, 90%, x = 24, 48 hours |
| Stochastic (S) model with annual (A) or hourly (H) TMR constraint with an x% renewable portfolio standard (RPS) constraint applied to non-$H_2$ demand | S-A-RPSx or S-H-RPSx<br>for x= 60, 70, 80 |
| Stochastic (S) model with an x% renewable portfolio standard (RPS) constraint applied to non-$H_2$ demand and no TMR constraint on H2 demand | S-A-RPSx_all for x= 60, 70, 80 |

The economic viability of the $H_2$ project is quantified using the LCOH, which is the annualized, all-in cost for the project developer per kg of $H_2$ produced. LCOH considers the following costs: a) fixed and variable costs of electrolyzer, b) fixed and variable cost of $H_2$ storage and compression, c) cost of electricity purchases (capacity and energy) to operate electrolyzer and compressor, d) fixed and variable costs of procured power generation and storage resources (referred to as power purchase agreement (PPA) resources from here on), and e) revenue obtained from sale of energy to the grid by PPA resources associated with the $H_2$ project. LCOH represents the lowest selling price per unit of $H_2$ that is required for the combined PPA and $H_2$ generation and storage assets to break even.

The robustness of capacity decisions under hourly TMR is assessed by exposing them to weather scenarios not considered during system sizing and measuring the extent to which they satisfy key operational constraints, including hourly supply-demand balance for $H_2$ and electricity and the TMR constraint (Eqs. 2-3). To do so, we fix the capacity decisions (investments and retirements) obtained from the deterministic and stochastic models and evaluate the optimal operation of the resulting electricity–$H_2$ system under alternative, out-of-sample weather scenarios. Robustness is quantified by the optimal value of slack variables in the key operational constraints that are subsequently penalized in the objective function to minimize constraint violation. The optimal value of the slack variables measures the degree to which in-sample capacity decisions fail to maintain operational feasibility under previously unseen weather scenarios. The optimal value of all slack variables in the CEM are zero by design.

We visualize the slack variable for the hourly TMR constrain in terms of: a) *number of hours where time-matching is unmet* and b) *share of hydrogen production that is unmatched in those hours.* The first metric corresponds to the number of hours in the year where the slack variable is non-zero for the TMR constraint (see S1.4). The second metric is the ratio of the sum of TMR slack variables (units of MWh) in hours with unmatched $H_2$ production divided by the total electricity demand for $H_2$ production in those hours.

To quantify the additionality of VRE resources added in the presence of a TMR constraint, we calculate the revenues per MW for PPA and grid VRE resources. For PPA resources, revenues accrue from the sale of energy to the grid as well as the TMR constraint, while for grid VRE resources, revenues accrue from the sale of energy and capacity to the grid and from contributing to the RPS constraint, if modeled.

# 3. Results

## 3.1. Deterministic vs. Stochastic model results

The optimal sizing of PPA resources, electrolyzer capacity, and $H_2$ storage for the deterministic models is highly sensitive to weather scenarios, as shown in Figure 2. This sensitivity is most pronounced with annual matching, where a single VRE resource typically dominates the PPA capacity mix depending on scenario-specific resource availability patterns. Under hourly matching, weather scenarios have less influence on PPA composition because meeting the more stringent constraint requires relying on both resources.

Electrolyzer and $H_2$ storage capacity decisions are more sensitive to weather scenario under hourly matching, where optimal electrolyzer capacity is between 1.4 to 2 times baseload $H_2$ demand, while $H_2$ storage capacity varies between 52-167 hours of exogeneous $H_2$ demand. We find that it is economical to oversize the electrolyzer relative to $H_2$ demand and deploy relatively less expensive $H_2$ storage rather than battery storage to satisfy hourly TMR, as in previous modeling studies.[12,16] This approach allows the electrolyzer to adjust its power consumption in response to hourly fluctuations in VRE generation from PPA resources, while $H_2$ storage buffers supply to meet the baseload $H_2$ demand. In contrast, under annual matching, operating the electrolyzer near baseload operation is generally more economical, as is lower $H_2$ storage capacity (2-7 hours of $H_2$ demand, Figure 2B) and electrolyzer capacity (1.1 times baseload $H_2$ demand, Figure 2C). The difference in electrolyzer dispatch between hourly and annual matching cases are also evident in the average hourly incremental generation profiles compared to the baseline (no-H2) model scenario. (Figure S1 vs. Figure S2).

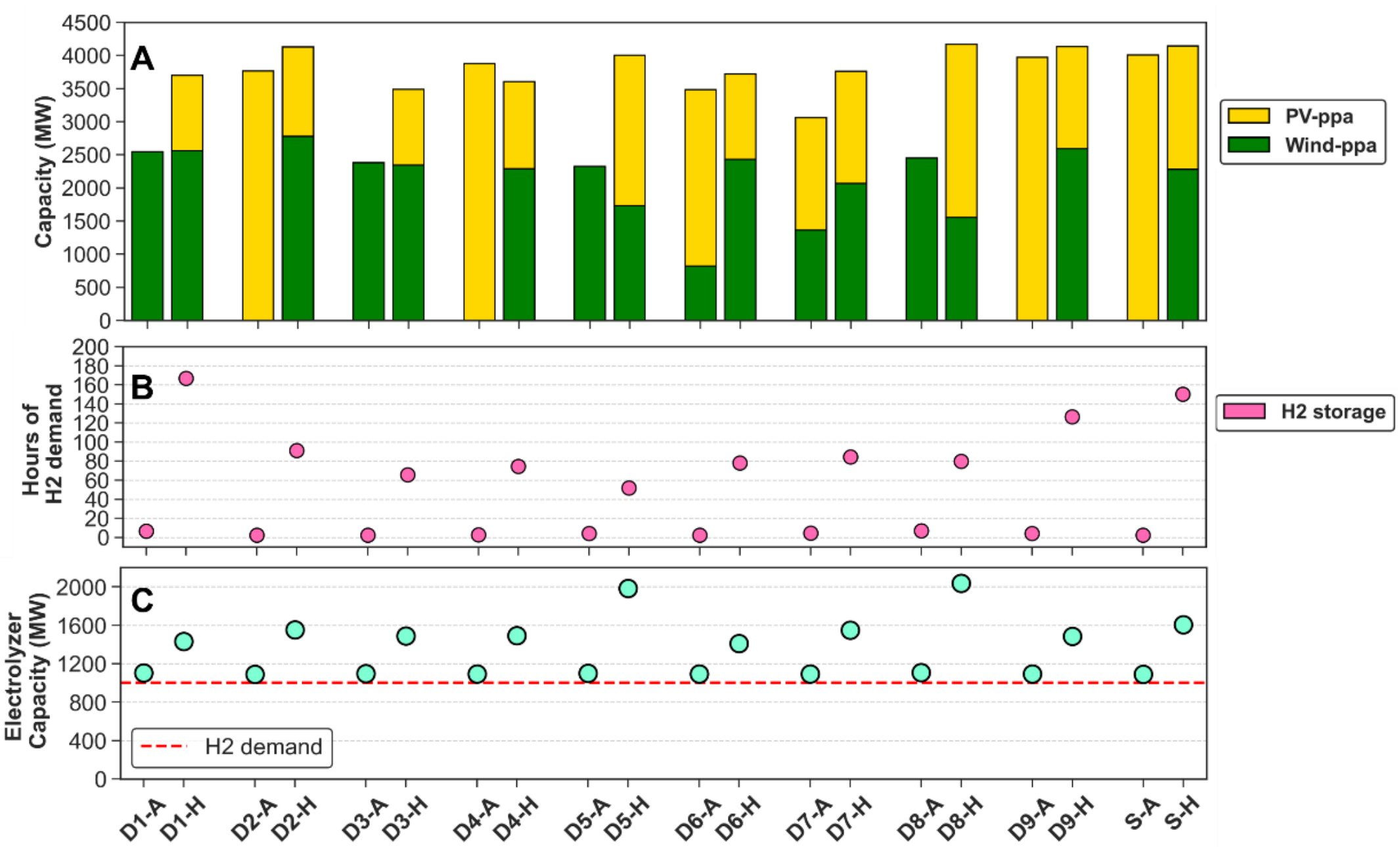

*Figure 2: PPA resources and $H_2$ project design for the stochastic and deterministic models under annual and hourly time matching requirements. PPA VRE and battery power capacity additions (A), energy storage capacity by storage technology (B), and installed electrolyzer capacity (C). The stochastic model (labeled "S-A" or "S-H," corresponding to annual or hourly TMR) co-optimizes design over nine weather scenarios, while deterministic model is solved for each of the nine weather scenarios (labeled "DX-A" or "DX-H" where X is weather scenario). $H_2$ storage capacity is reported in terms of hours of $H_2$ demand, which is calculated by dividing the $H_2$ storage capacity by the baseload $H_2$ demand (18.4 tonnes/hour). For batteries,*

*energy storage capacity in terms of hours of $H_2$ demand is calculated by dividing the battery energy capacity (in GWh) by the electricity required for the electrolyzer to produce one hour of $H_2$ demand (i.e., 1GWh / 18.4 t $H_2$) — e.g., a 2 GWh battery is storage is equivalent to two hours of $H_2$ demand. No PPA battery storage is deployed across modeled scenarios.*

The stochastic model preserves the key differences in asset sizing between annual and hourly matching observed in the deterministic model. Under hourly matching, the stochastic model favors deploying more $H_2$ storage, oversizing the electrolyzer, and combining wind and solar PPA capacity. Notably, the stochastic model's optimal $H_2$ storage and electrolyzer capacities approach the highest values observed across individual deterministic weather scenarios. This suggests that increased $H_2$ storage and electrolyzer capacity, which collectively represent a form of electricity demand flexibility, hedge against both intra-annual and inter-annual weather variability in satisfying the hourly TMR constraint.

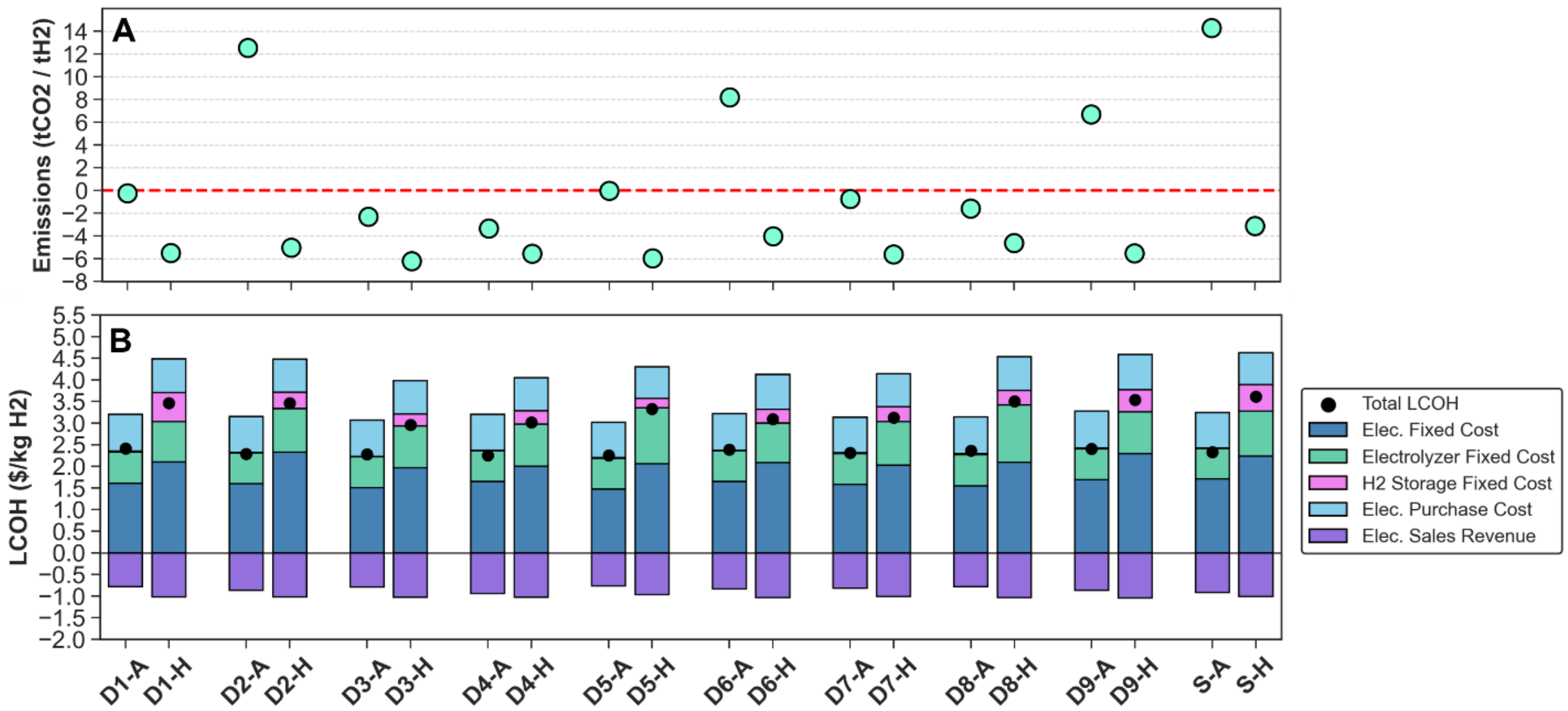

*Figure 3: A) Consequential emissions intensity of $H_2$ production under different model cases, corresponding to different TMR (annual, hourly) and model configurations – stochastic and deterministic under nine different weather scenarios. B) Levelized cost of $H_2$ production associated with deterministic and stochastic model runs under annual and hourly TMR. Model configuration nomenclature as defined in Table 1 and Figure 1 caption.*

Figure 3A summarizes the emissions impact of electricity-based $H_2$ production under annual versus hourly matching. Under annual matching, consequential emissions span a wide range — from near-zero to 14 $tCO_2/tH_2$. In contrast, hourly matching consistently yields negative consequently emissions, indicating that a grid operating with $H_2$ demand and a hourly TMR constraint produces lower emissions than the baseline. For context, attributional life cycle greenhouse gas emissions from natural gas-based $H_2$ production without $CO_2$ capture are approximately 10 $tCO_2eq/tH_2$[34,35]. Among the nine deterministic model cases with annual matching, six achieve near-zero or negative emissions. It should be noted, however, that hourly time-matching requirements, by necessitating larger electrolyzer and battery capacities, may increase embodied manufacturing emissions not captured in our emissions metric, potentially offsetting some of the grid-level emissions benefits documented here.

The emission outcomes seen in Figure 3A result from changes in capacity deployment and utilization induced by the TMR, summarized in Figure S4. Under annual matching, positive emissions occur when solar dominates the PPA capacity mix. This is because PPA solar displace grid solar deployed in the no-H2 baseline case, and because during hours when the PPA solar is not generating additional gas generation is needed (Figure S1) to meet $H_2$ production's near-baseload electricity demand, given the limited energy storage capacity (Figure 1B). Conversely, grid VRE and gas generation are largely unaffected when sufficient wind capacity is deployed to meet the annual matching constraint, resulting in near-zero emissions impact. The more stringent hourly matching constraint leads to a mix of wind and solar capacity and flexible electrolyzer operation, all of which mitigate the mismatch between incremental supply and demand without inducing additional gas generation. Moreover, the greater build out of wind and solar capacity relative to incremental electricity demand under hourly matching (Figure 1A) leads to higher (6-13%) VRE curtailment as compared to near-zero curtailment in the annual matching cases (Figure S3).

These findings align with prior assessments indicating that annual matching with grid-PPA competition can yield higher emissions than hourly matching.[12,13,16] However, here we demonstrate that generalizing from deterministic assessments requires caution, as emissions outcomes under annual matching are highly sensitive to the selected weather scenario. This underscores the need to consider inter-annual weather variation in assessment of long-term emissions impacts of clean electricity procurement strategies based on hourly matching.

Figure 3A shows that, compared to deterministic model outcomes across weather scenarios, the stochastic model estimates higher emissions under annual matching and smaller emission reductions under hourly matching. Regarding resource deployment, the stochastic model retains some coal capacity under both TMR approaches, primarily to meet the resource adequacy requirement (Eq. S1), whereas deterministic cases virtually eliminate all coal capacity.

The lower emissions impact of hourly matching versus annual matching comes with a cost premium (Figure 3B), that ranges between \$0.68-1.18/kg $H_2$ across the different weather scenarios with the deterministic model and \$1.29/kg $H_2$ under the stochastic model. This highlights how a deterministic model may underestimate the cost premium associated with hourly matching. For context, the LCOH of $H_2$ production via natural gas based $H_2$ production in the U.S. context, with and without $CO_2$ capture is estimated to be around \$1/kg and \$1.5-2 /kg, respectively.[34–36] Hourly matching is more expensive because it requires more PPA VRE resources, as well as greater electrolyzer and $H_2$ storage capacity (Figure 2).

### 3.2 Out-of-Sample Analysis

Figure 4 shows that stochastic modeling produces a solution that is more robust at meeting an hourly TMR when exposed to out-of-sample weather scenarios than the deterministic model. Of the 10 out-of-sample weather scenarios we tested, the stochastic model solution is able to accommodate all without having to relax the hourly TMR constraint, whereas the deterministic model design solutions require relaxing the constraint for 48 of the 90 runs (9 deterministic cases x 10 out-of-sample weather scenarios per case). Across the cases, the number of hours in which the hourly time-matching requirement was not fully satisfied ranges from 0 to 744. The average share of electricity demand for $H_2$ production that is not matching during those hours is calculated for each case, with a median of 16% across cases and a maximum of 51%.

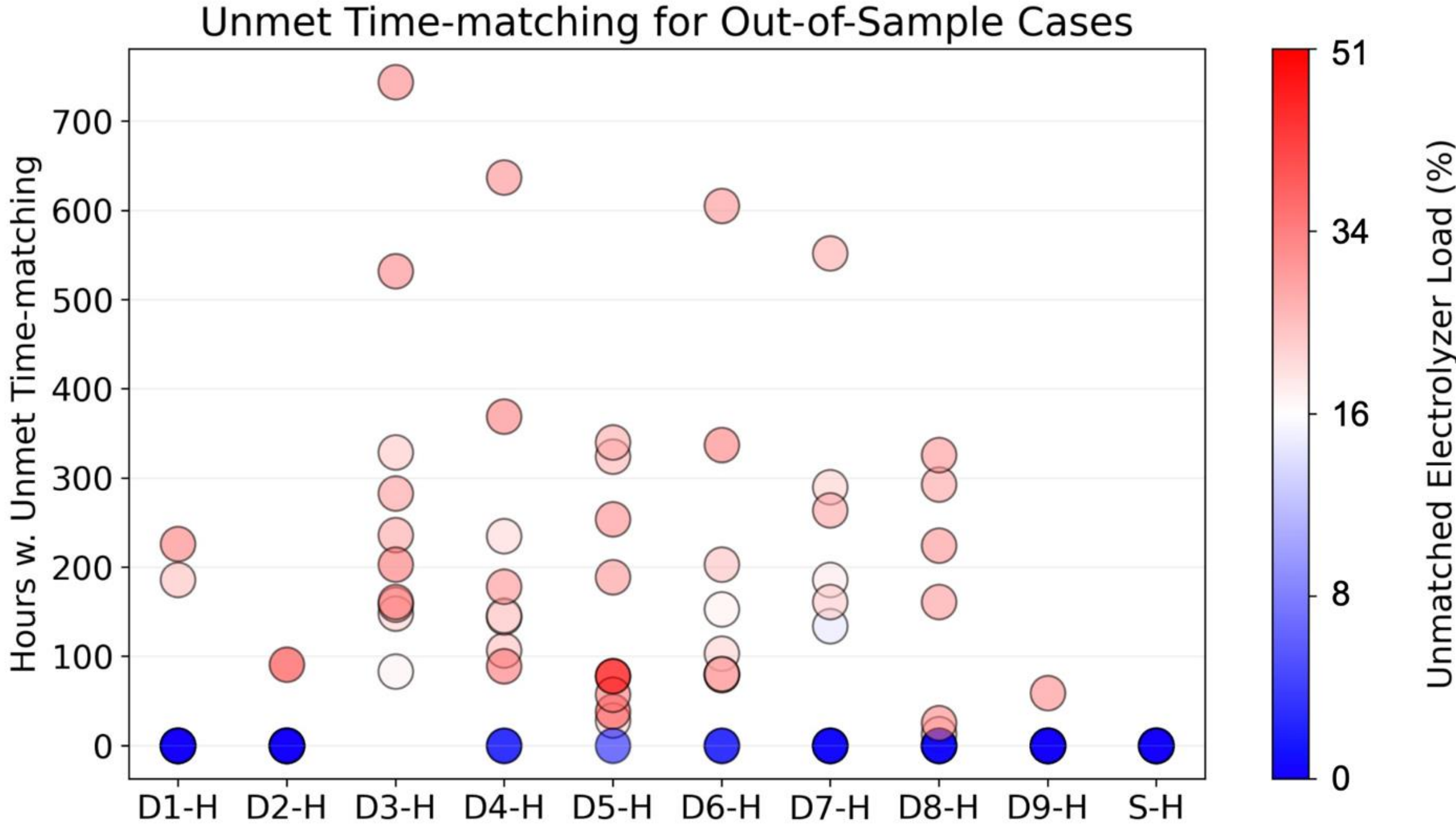

*Figure 4: Unmet time-matching for out-of-sample cases for stochastic and deterministic model design solutions under an hourly time-matching requirement. Marker position corresponds to the left-axis, which shows the number of hours in an out-of-sample case in which the full time-matching requirement was not fully satisfied. Unmet time-matching is enabled by utilizing a slack variable that permits electricity used by the electrolyzer to be greater than the generated electricity by PPA resources in that hour. The color of the markers indicates the average share of the time-matching requirement that was not met during hours when the requirement was not fully satisfied (i.e., 'Unmatched Electrolyzer Load (%)'), which is calculated as the sum of TMR slack variables (corresponding to MWh) in hours with unmatched $H_2$ production divided by the total electricity demand for $H_2$ production in those hours.*

The robustness of the stochastic solution to out-of-sample weather scenarios is in part driven by deployment of hydrogen storage (Figure 2); however, this level of $H_2$ storage build-out may exceed real-world availability (see next section). We replicated the out-of-sample analysis described above for stochastic cases where $H_2$ storage buildout is constrained (Figure 5) to test whether the stochastic solution remains robust when $H_2$ storage is limited. Somewhat surprisingly, we find that the stochastic model solutions remain robust to out-of-sample weather years, with no slack utilization across the 90 cases (10 out-of-sample weather scenarios times the 9 cases described in the following section). This suggests that increasing the capacity of wind and solar resources in a balanced portfolio maintained the robustness of the solution compensating for constraints on $H_2$ storage capacity.

### 3.3 Impact of changing compliance on TMR constraint and $H_2$ storage capacity

We use the stochastic model to investigate two practical considerations for implementing hourly matching in electricity-based $H_2$ production. First, we explore the impact of reducing the compliance threshold (defined by $\alpha_{TMR}$ in Eq. 2) for hourly matching to slightly less than 100% (80% or 90%), because strict hourly matching may be prohibitively expensive for some applications. Second, we limit the capacity of above-ground $H_2$ storage, which may result from implementation constraints (e.g., land use[37]) arising from the low volumetric energy density and flammability of hydrogen. For reference, the amount of $H_2$ storage capacity installed for the stochastic model with hourly TMR (2,769 tonnes – see Figure 1B) is similar to the size of the largest operational underground $H_2$ storage facility[38].

As expected, reducing compliance to 80% or 90% decreases PPA and energy storage capacity investments (Figure 5A,E) and, consequently, the cost of $H_2$ production (Figure 5D). However, the cost savings are relatively modest—6.8% and 3.4% for 80% and 90% compliance, respectively (Figure 5D)—when compared to the 35% cost increase of hourly versus annual matching in the stochastic model. The reduced PPA capacity under partial compliance also diminishes system-wide emissions reductions, with 80% compliance achieving near-zero emissions impact rather than negative emissions (Figure 5C). These results have implications for procurement strategy: if consumers prioritizing hourly matching seek primarily to minimize their own emissions impact rather than achieve system-wide reductions, then partial compliance (e.g., 80% in the modeled case) may offer a more practical and cost-effective approach than hourly matching with 100% compliance.

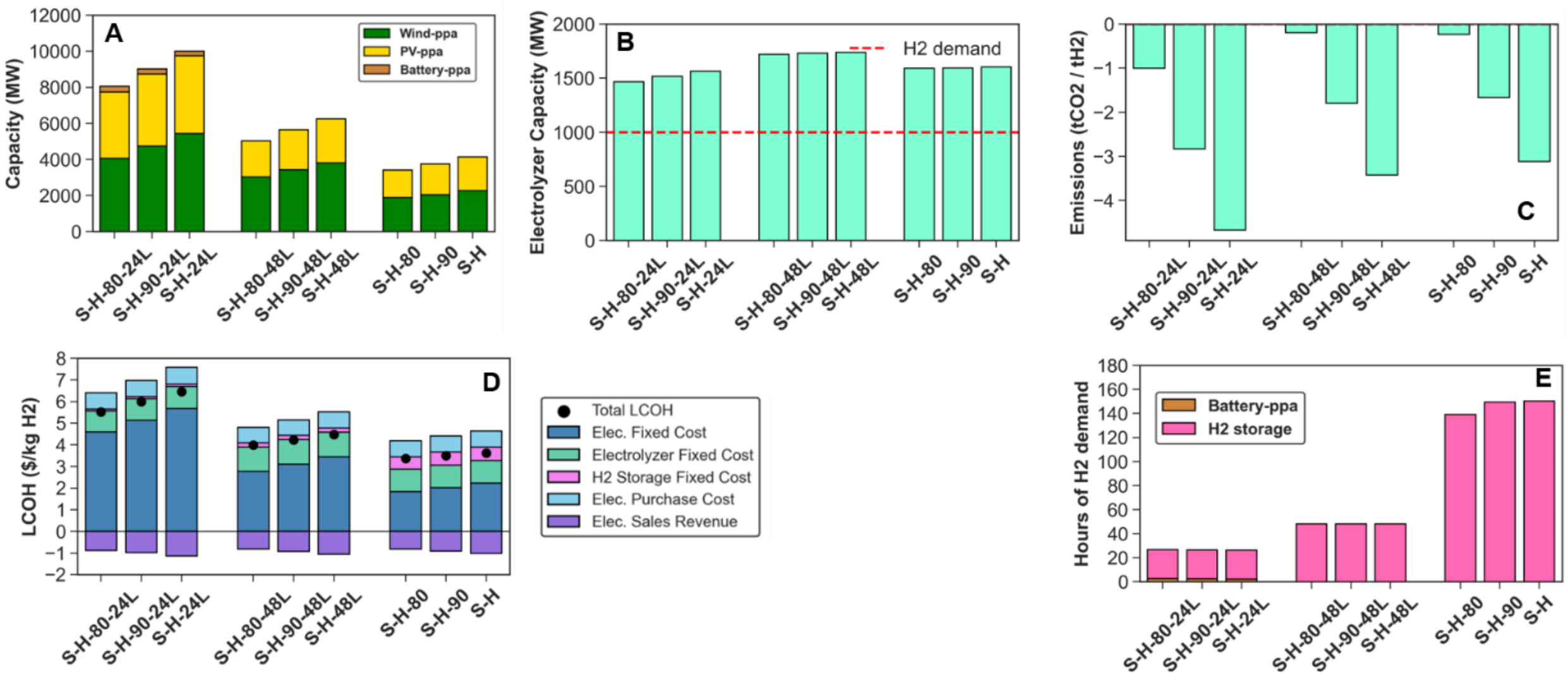

*Figure 5: Impact of limiting $H_2$ storage capacity and degree of compliance with the hourly matching constraint on the stochastic model outcomes: A) PPA capacity mix, B) electrolyzer capacity, C) consequential emissions, D) levelized cost of hydrogen (LCOH) and E) energy storage capacity. Results based on stochastic model with hourly time matching requirement. S-H = Stochastic model with total hourly matching; S-H-X = Stochastic model with hourly matching with X percent (80,90) compliance of the constraint; S-H-X-YL = Stochastic model with hourly matching with X percent compliance and limit on H2 storage capacity equal to Y hours of $H_2$ demand (Y = 24, 48). S-H-YL= Stochastic model with hourly matching with 100 percent compliance and limit on $H_2$ storage capacity equal to Y hours of $H_2$ demand (Y = 24, 48). See Table 1 for further explanation of scenario labels.*

Compared to varying compliance thresholds, restricting $H_2$ storage availability has a more substantial impact on PPA and $H_2$ asset design under hourly matching (Figure 5). When $H_2$ storage is constrained to 24 or 48 hours of rated $H_2$ demand, achieving 100% hourly matching compliance requires substantially more PPA VRE capacity—approximately 50% and 150% increases (Figure 5A), respectively—along with additional electrolyzer capacity and battery energy storage in some cases. These capacity additions translate to LCOH increases of 78% and 24% relative to the unconstrained storage case (Figure 5D) for the 24 and 48-hour storage limit scenarios, respectively. In addition, relaxing the compliance threshold offers greater cost savings when storage availability is limited. For example, reducing hourly matching compliance from 100% to 90% with a 24-hour $H_2$ storage limit lowers costs by 7% as compared to 3.3% under the unconstrained storage availability case. Finally, limited $H_2$ storage availability also reduces the degree of compliance needed to achieve near-zero emissions impact (Figure 5C).

### 3.4 Impact of an RPS Policy

Projects subject to temporal matching requirements will not exist in isolation but within existing policy frameworks governing electricity system transitions. Here, we examine the interaction between TMR for new electricity-based $H_2$ production and RPS constraints on existing electricity demand, which is a common decarbonization policy employed across many regions in the U.S. and around the world.

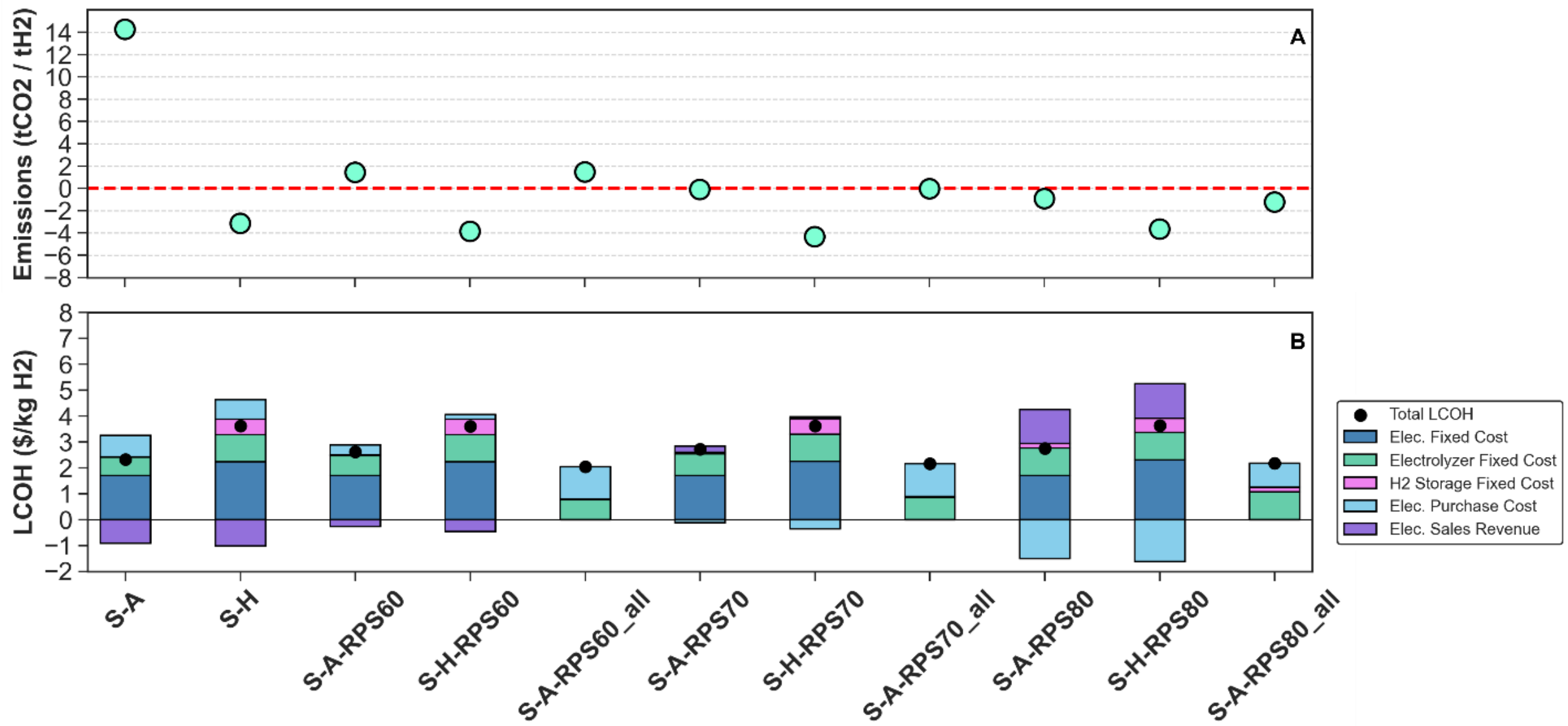

*Figure 6: Consequential emissions (A) and levelized cost of hydrogen (B) from stochastic model runs with varying renewable portfolio standard (RPS) requirements for non-$H_2$ electricity demand (no RPS, 60%,70% and 80%). "RPS 60" indicates 60% minimum annual VRE generation for non-$H_2$ electricity demand and so on. "RPSX_all" includes electrolyzer consumption in the RPS constraint and without a time-matching requirement. S-A/S-H: annual/hourly time-matching. S-A-RPSXX/S-H-RPSXX: annual/hourly time-matching for electrolyzer with XX% RPS for non-$H_2$ demand. Note: For "RPSX_all" scenarios, electricity purchase cost by electrolyzer and compressor includes :a) energy price, b) capacity price and c) RPS price. Prices are summarized in Figure S6. See Table 1 for further explanation of scenario labels.*

A binding RPS constraint on non-$H_2$ electricity demand mitigates the displacement of grid VRE capacity that would otherwise occur when adopting new demand with an annual TMR (compare Figure S4 to Figure 3). In effect, an RPS constraint reduces competition between VRE capacity deployed to meet TMR constraint and VRE capacity deployed solely based on economics of electricity supply to the grid. Consequently, the emissions impacts of annual time-matching under a binding RPS constraint for non-$H_2$ electricity demand are low or negative and largely similar to that of hourly matching. An RPS constraint has minimal impact on emissions under an hourly requirement, which does not lead to displacement of grid

VRE capacity even in the absence of an RPS constraint (Figure S5). RPS requirements have minimal impact on $H_2$ asset deployment (Figure S6), with no change under hourly matching and only slight increases in electrolyzer and $H_2$ storage capacity under annual matching to capitalize on increased electricity price volatility from higher VRE penetration.

A binding RPS requirement on non-$H_2$ electricity demand suppresses energy prices—specifically, the shadow price of the hourly electricity supply-demand balance constraint (Figure S7)—which affects LCOH through two countervailing mechanisms: (a) it reduces electricity purchase costs for $H_2$ production and storage, and (b) it reduces revenue from electricity sales by PPA VRE resources. These opposing effects largely offset each other, resulting in minimal net impact on LCOH as seen in Figure 6B.

However, the impact of an RPS requirement on TMR prices —referring to the shadow price of the TMR constraint is more substantial, particularly under annual matching (Figure S7). TMR prices can be interpreted as the cost of energy attribute certificates (EACs) to be purchased by the $H_2$ producer from the PPA VRE generators as part of their executed PPA. TMR prices increase under binding RPS constraints to compensate PPA VRE resources for reduced energy revenues caused by wholesale energy price suppression. Under annual matching, TMR prices increase from \$14/MWh without an RPS to \$26/MWh, \$36/MWh, and \$55/MWh under RPS60, RPS70, and RPS80, respectively.

We also evaluate an alternative scenario where $H_2$ producer's electricity demand is included within the binding RPS constraint rather than subject to separate matching requirements. As shown in Figure 6, this approach yields near-zero consequential emissions outcomes but at lower cost to the $H_2$ producer compared to annual and hourly time-matching. The cost advantage arises because $H_2$ producers only need to contract for VRE supply equal to the RPS requirement (less than 100%), rather than match 100% of their consumption. For example, under RPS60, a $H_2$ producer consuming 54.3 kWh/kg $H_2$ must procure EACs associated with the RPS for only 60% of electricity consumption or 32.58 kWh/kg $H_2$. With an RPS price of \$25.7/MWh (shadow price of RPS constraint, reported in Figure S7), this translates into a cost of \$0.84/kg $H_2$. In contrast, the annual TMR price under RPS60 is approximately \$26/MWh (Figure S7), resulting in a TMR EAC cost of \$1.4/kg $H_2$ (54.3 kWh/kg x \$26/MWh)—67% higher because annual matching requires EACs for 100% of consumption.

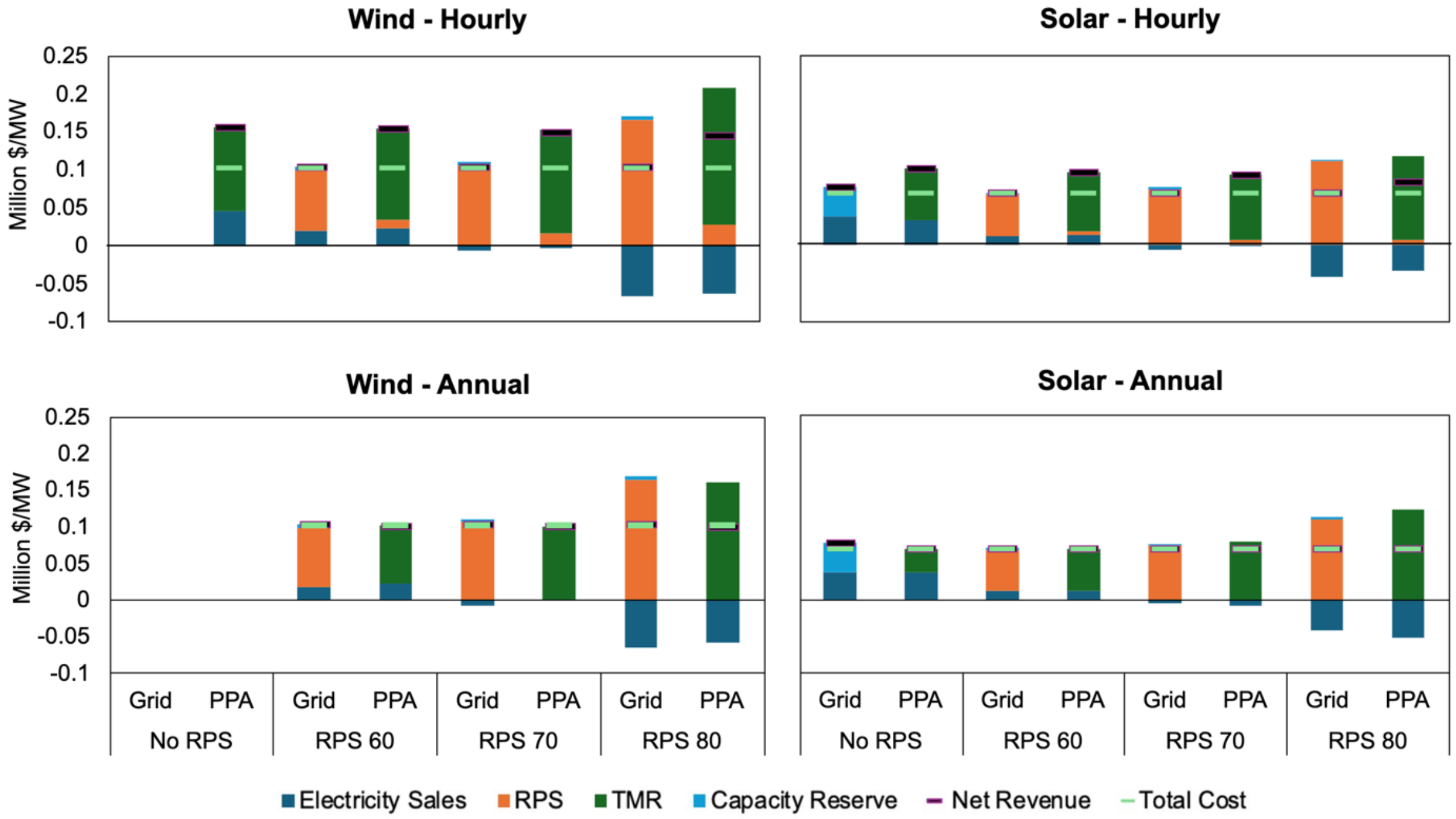

*Figure 7: Comparison of cost and revenues for newly installed grid and PPA VRE resources under different scenarios with and without an RPS requirement on non-$H_2$ electricity demand. 'Grid' resources are new VRE that serve the grid generally, whereas 'PPA' resources are those built to meet the time-matching requirement of the $H_2$ project. 'Electricity sales' is the revenue earned from grid injections by both PPA and grid resources. 'RPS' revenues are earned by only grid resources for selling RPS energy attribute certificates. 'TMR' revenues are earned only by PPA resources for satisfying the time-matching requirement. 'Capacity Reserve' revenues are earned only by grid resources for contributing to capacity reserve requirement (Eq. S1). 'Net Revenue' sums all revenue sources. 'Total Cost', reflects the investment and operating cost of VRE capacity as per cost assumptions summarized in Table S2.*

Figure 7 shows revenue sources for grid and PPA VRE resources across several RPS requirements (60%, 70%, and 80%) and a no-RPS scenario, expressed on a dollar-per-MW basis. Here, revenues from the various constraints are estimated based on the corresponding shadow prices, summarized in Figure S7-S8. Under the modeled CEM framework, revenues are necessarily greater than or equal to costs for all installed resources,[39,40] which is substantiated in Figure 7. Under hourly time-matching, PPA resources derive the majority of their revenue from the TMR constraint across all scenarios, reflecting a strong causal link between the TMR constraint and resource deployment and implying strong additionality. Under annual time-matching without an RPS constraint on non-$H_2$ demand, PPA resources (solar) earn a smaller share of revenue from the TMR, suggesting that resource deployment is less strongly tied to the constraint, indicating weaker additionality relative to hourly time-matching. When a binding RPS constraint on non-$H_2$ demand is introduced, however, annual time-matching achieves comparably strong additionality—PPA resource revenue is predominantly sourced from the TMR constraint, similar to the pattern observed under hourly time-matching. This suggests that annual TMR coupled with a binding RPS policy on non-$H_2$ demand offers similar levels of additionality as hourly time-matching but at a lower cost to the $H_2$ producer (Figure 6B).

Under a binding RPS constraint, grid resources earn the majority of their revenue from RPS payments. Without an RPS, their revenue comes from a mix of capacity reserve payments and electricity sales. Electricity sales comprise a minor share of revenues for all VRE resources. As the stringency of the RPS constraint increases, electricity sales generate increasingly negative revenues because renewable resources must generate during periods with negative electricity prices to comply with both RPS and TMR obligations.  Capacity reserve payments constitute a minor share of the revenue stack for grid resources under a binding RPS, reflecting the declining capacity value of VRE resources with increasing deployment.[41]

The revenue of PPA resources under an hourly time-matching requirement exceeds their costs due to the cap on excess energy sales to the grid (Eq. 4). This additional "rent" earned by PPA resources indicates that, without the excess energy sales cap, deploying more PPA resources would be profitable. This explanation is corroborated by the observation that PPA resources do not earn rent under the annual time-matching requirement, where excess energy sales are prohibited (Eq. 3). As noted by others, the excess energy sales cap constraint for hourly TMR is a modeling construct that allows for distinguishing between resources deployed to meet the TMR constraint and those deployed based purely on power grid economics.[16,25]

## 4. Discussion

We investigated the impact of inter-annual weather variability on the composition, costs, and emissions of electrolyzer projects subject to annual and hourly time-matching requirements using a CEM that accounts for competition between VRE resources for $H_2$ production and grid applications. We also tested the impacts of (a) relaxing the hourly TMR constraint (b) limiting $H_2$ storage availability and the implications for $H_2$-related electricity demand flexibility, and (c) enforcing an RPS requirement.

Our results reaffirm prior findings that annual TMR yields lower cost but higher emissions than hourly TMR[5,12,13,16,25] and additionally show that annual time-matching outcomes are highly sensitive to the assumed weather scenario. This underscores the importance of incorporating inter-annual weather variability – for example, through the stochastic modeling approach evaluated here - when assessing system design and long-term emissions impacts of alternative TMRs. The stochastic model preserves the qualitative differences in system design, costs, and emissions impacts between annual and hourly matching observed in deterministic approaches, while better accommodating inter-annual VRE variations. This is evidenced by stochastic designs yielding zero periods of unmatched $H_2$ production in the out-of-sample dispatch analysis. However, the difference in cost between hourly and annual TMR is greater in the stochastic model, suggesting that deterministic approaches may underestimate the cost premium of hourly matching, while also likely generating solutions that could fail to match the electrolyzer's consumption with procured clean electricity in practice.

We also examined practical implementation considerations for hourly matching, including partial compliance (matching less than 100% of hourly demand) and limited availability of $H_2$ storage that enables flexible operation. Partial compliance may appeal to consumers seeking near-zero rather than deeply negative consequential emissions at reduced cost, particularly when $H_2$ storage is constrained. Achieving full compliance would likely require spot markets for hourly time-matched EACs, where consumers can purchase shortfalls not covered by their long-term procurement contracts. Such markets could exhibit highly skewed pricing due to the coincidental nature of regional VRE supply: a few high-price hours during periods of limited VRE availability, accompanied by many low-price hours when VRE is abundant. Figure S8 illustrates this distribution of hourly TMR prices across various stochastic model runs. This finding is consistent with empirical evidence from the European EAC market, where Scholta and Blaschke[42] document increased instances of VRE demand exceeding VRE supply during nighttime hours compared to day time hours.

Limited $H_2$ storage availability, which constrains flexible electrolyzer operation, increases $H_2$ production costs by 24–78% across the evaluated scenarios. More broadly, this underscores the critical role of demand flexibility in reducing VRE capacity requirements and costs under hourly matching. For baseload demands such as current data center operations[43,44], hourly matching may prove impractical without measures to improve demand flexibility.

Binding RPS requirements on existing electricity demand substantially diminish the incremental emissions benefits of hourly versus annual matching, as RPS constraints provide dedicated revenue streams for grid-oriented VRE deployment that effectively decouple it from competition with PPA VRE capacity. This reduced competition mitigates incremental use of fossil-generation seen with annual matching without any RPS constraints on non-$H_2$ electricity demands. In addition, we show that incorporating the new electricity demand directly into binding RPS constraints—rather than imposing separate time-matching requirements—achieves near-zero consequential emissions outcomes but at a lower overall cost to the consumer. This finding suggests that in regions with binding renewable energy targets or equivalent grid decarbonization policies, temporal matching-based clean electricity procurement may be less attractive as an emissions reduction strategy as compared to participating in existing regulation-driven markets.

Future research could expand the stochastic modeling framework employed here to assess how transmission constraints and restrictions on the spatial boundary for VRE procurement impact PPA system design, cost, and emissions outcomes of hourly versus annual matching. Prior work using deterministic demand-centric modeling approaches suggests that transmission constraints could affect grid dispatch and therefore the emissions impacts of hourly matching[11]. Our analysis also did not consider spatial heterogeneity in VRE resource quality and grid integration costs, which could influence the relative competitiveness of different resources for voluntary clean electricity procurement. Another important aspect not evaluated here is how increasing adoption of time-matching as a clean electricity procurement strategy affects the cost of electricity supply to other consumers. The stochastic modeling framework employed here can also be extended to quantify regional variations in the impacts of TMR-based clean electricity procurement strategies. For instance, a recent study employing a single weather-scenario CEM framework found substantial regional variation in the emissions and cost impacts of hourly time-matching based VRE procurement[26].

Although this study demonstrates the importance of accounting for inter-annual weather variability using historical data, future work should examine how projected changes in VRE resource availability as a result of climate change and long-term system evolution — including the multi-year build-out of hydrogen infrastructure — affect the cost and emissions outcomes of alternative clean electricity procurement strategies. Finally, while our study has examined system impacts of voluntary procurement by a single consumer type, real-world markets involve consumer with varying demand patterns, flexibility potential, and preferences for clean electricity procurement. Decentralized modeling approaches that capture individual consumer consumption patterns, procurement decisions and risk preferences could therefore provide insights into likely market outcomes under large-scale adoption of voluntary clean electricity procurement strategies.

**Supporting information.** Additional results for evaluated scenarios; summary of modeling approach and input assumptions and implementation details.

## 6. Acknowledgements

An initial version of this work was included as part of MAG's Masters thesis research at MIT, where he was supported by funding from the Future Energy Systems Center at the MIT Energy Initiative.

**Supporting information for**

# Emissions and cost tradeoffs of time-matched clean electricity procurement under inter-annual weather variability – case study of hydrogen production

Michael Giovanniello[1], Dharik S. Mallapragada[2]*

1. MIT Energy Initiative, Massachusetts Institute of Technology, Cambridge, MA 02139

2. Chemical and Biomolecular Engineering Department, Tandon School of Engineering, New York University, Brooklyn, NY 11201

*Correspondence: Dharik S. Mallapragada

**Email**: dharik.mallapragada@nyu.edu

**Summary**: This supporting information contains 19 pages, 11 figures, 6 tables, and 2 equations.

# S1. Modeling approach

## S1.1. Capacity expansion model overview

Table S 1 summarizes key constraints in the capacity expansion model for system-level and project-specific operation and their relation to investment decisions.

*Table S 1. Summary of constraints associated with system and technology operation in the DOLPHYN CEM used in this study[23,24]. Further description of the specific policy constraints of interest to the model are provided in the supplementary information section S1.1. PPA = Power Purchase Agreement. "PPA generators" refers to those generators contracted by $H_2$ production to satisfy the time-matching requirement constraint.*

| Constraint Scope | Constraint Type |
|---|---|
| System-level constraints | • Supply-demand balance (hourly)<br>• Capacity reserve margin constraint ensuring resource adequacy (hourly)<br>• Renewable energy share requirements (annual)<br>• Time-matching requirements (TMR) associated with PPA generators to serve electricity demand for H2 production (hourly or annual) |
| Thermal generators | • Power output must be less than available capacity at each time step<br>• Minimum stable power output for each time step<br>• Linearized unit commitment<br>• Ramp rate limits (up & down) |
| Renewable Energy Generators (VRE, Hydro) | • Capacity constraint with time-dependent availability factor at each time step<br>• Minimum production rate (Hydro only) at each time step<br>• Ramp rate limits (Hydro only) at each time step |
| Battery energy storage | • Power and energy capacity constraints at each time step<br>• Storage inventory balance constraints with charging/discharging efficiency<br>• Minimum amount of energy in storage in at each time step |
| Electrolyzer | • Capacity constraint with constant availability factor at each time step<br>• Power consumption by electrolyzer associated with $H_2$ production in each time step (sector-coupling)<br>• Minimum $H_2$ production rate at each time step<br>• Ramping rate limits at each time step |
| $H_2$ storage | • Charging power (compression) and energy capacity constraints at each time step<br>• Storage inventory balance constraints with charging/discharging efficiency at each time step<br>• Minimum amount of energy in storage at each time step<br>• Maximum energy storage capacity limit (implemented for some of the evaluated scenarios) |

## S1.2 Key system-level constraints

Aside from the supply-demand balance at each time step for electricity and $H_2$ commodities and constraints related to time-matching requirement (TMR) described in the main text, the model includes the following system level constraints.

- **Resource adequacy constraint (Eq. S1)**: this constraint, referred sometimes as the capacity reserve margin, enforces the need to procure "firm" generation capacity in excess of demand ($\delta_t$) by the specified amount (i.e. reserve margin) in each hour of the year. Here the firm capacity contribution of each resource, after accounting for a derating factor($\eta$), is calculated based on: a) the installed capacity ($Y_g$) in case of thermal plants ($g \in D$), b) the hourly available generation ($X^a_{g,t}$) in case of non-dispatchable resources like renewables and c) the difference between discharging and charging rates ($X^d_{s,t} - X^c_{s,t}$) in case of energy storage and d) the negative of net consumption ($gen^{ELY}_{r,t}\beta_r$) in case of flexible demand (e.g. electrolyzers). Table S5 reports the derating factor assumptions for each resource as well as the enforced capacity reserve margin ($\alpha_{CRM}$).

$$\sum_{g \in D} Y_g \eta_g + \sum_{g \in ND} X^a_{g,t} \eta_g - \sum_{s \in S} \left( X^d_{s,t} - X^c_{s,t} \right) \eta_s - \eta^{ELY} gen^{ELY}_t \beta^{ELY} \geq (1 + \alpha_{CRM}) \delta_t \qquad \text{(S1)}$$

- **Renewable Portfolio Standard (RPS) requirement (Eq. S2)**: this constraint requires that annual generation from existing and new VRE resources serving the grid and not contracted with the electrolyzer (i.e. so-called 'grid' or 'non-PPA' resources, $g \in VRE_g \backslash TMR_g$), must be at least equal to a pre-specified ($\kappa$) of annual electricity demand ($\delta_t$). Note that electricity demand does not include electricity consumed for $H_2$ production.

$$\sum_{g \in VRE_g \backslash TMR_g} \sum_{t \in T} gen_{g,t} \geq \kappa \times \sum_{t \in T} \delta_t \qquad \text{(S2)}$$

## S1.3. Out-of-sample analysis

The out-of-sample analysis takes the capacity decisions obtained from the solution of the stochastic and deterministic model (the *design* models) and tests their performance using VRE availability for other (*out-of-sample model)* VRE scenarios, corresponding to different weather scenarios. The purpose is to assess the robustness of solutions generated by the stochastic and deterministic model, as well as gain insights into possible contract designs for procurement of hourly renewable electricity and need for real-time markets to balance supply and demand for clean energy attribute certificates (EACs). The cost-optimal system design for the power grid and $H_2$ production and storage obtained by the *design* model are fixed in the *out-of-sample dispatch* model where the operation of this system is optimized using an *out-of-sample* weather scenario instead of the VRE weather used to generate the design solution.

Without some level of flexibility in the hourly-time matching constraint, the out-of-sample dispatch model may be infeasible. To maintain model feasibility and quantify how much operational flexibility is required, a slack term, $TMR_{slack,t}$, is introduced into the hourly TMR constraint (Eq. 2) for the out-of-sample model runs. This slack term enables the electrolyzer to operate with imperfect matching from contracted resources. Without this slack term, the model would be infeasible whenever contracted VRE resources are unable to meet the TMR constraint. Utilization of $TMR_{slack,t}$ is penalized at $500/MWh in the objective function, which is lower than the cost of not serving the grid ($9000/MWh – See Table S4) but well above the cost of the most expensive electricity generator. This ensures that the slack variable is only used when electricity from the PPA resources is scarce, but that the time matching constraint will not take priority over serving grid demand during grid scarcity events. For the out-of-sample analysis, we focus on the impact of the more stringent hourly-time-matching requirement since this is the more stringent of the two TMR constraints, and because hourly TMR is being considered as part of multiple regulatory and voluntary efforts for characterizing electricity-related emissions of end-users.

## S1.4. Modeling Renewable Portfolio Standards (RPS) and Time-Matching Requirements in the stochastic model

Under the stochastic model, the annual RPS requirement is enforced as a separate constraint for each of the nine weather scenarios. Practically, this means that the resource mix reflected in the model solution must be sufficient to meet the RPS requirement in every weather scenario, and is therefore sized to meet the most challenging weather scenario(s). The consequence is that the annual VRE generation under some scenarios will exceeds the RPS requirement in most scenarios and be binding in a subset of scenarios. Effectively, the shadow price of the RPS requirement constraint in many of the modeled weather scenarios is likely to be zero.

In addition, we do not count generation from PPA resources towards meeting the RPS requirement so as to avoid double counting the clean energy attributes of the resource. In addition, PPA VRE resources are disallowed from contributing to the resource adequacy requirement (see Eq. S1). This ensures that resources designated as PPA are being built purely to serve the TMR requirement constraint associated with the electrolyzer demand, as opposed to being built to serve either the RPS or resource adequacy constraint.

## S1.5 Implementation details

The numerical experiments were run on high-performance computing clusters available to authors at MIT and NYU using Gurobi 11.0.2. For the resulting linear problems, we used the interior point method without crossover, a barrier convergence tolerance of $10^{-8}$ and 32 threads per model. The maximum memory requirement for any individual model run was 102 GB.

# S2. Additional results

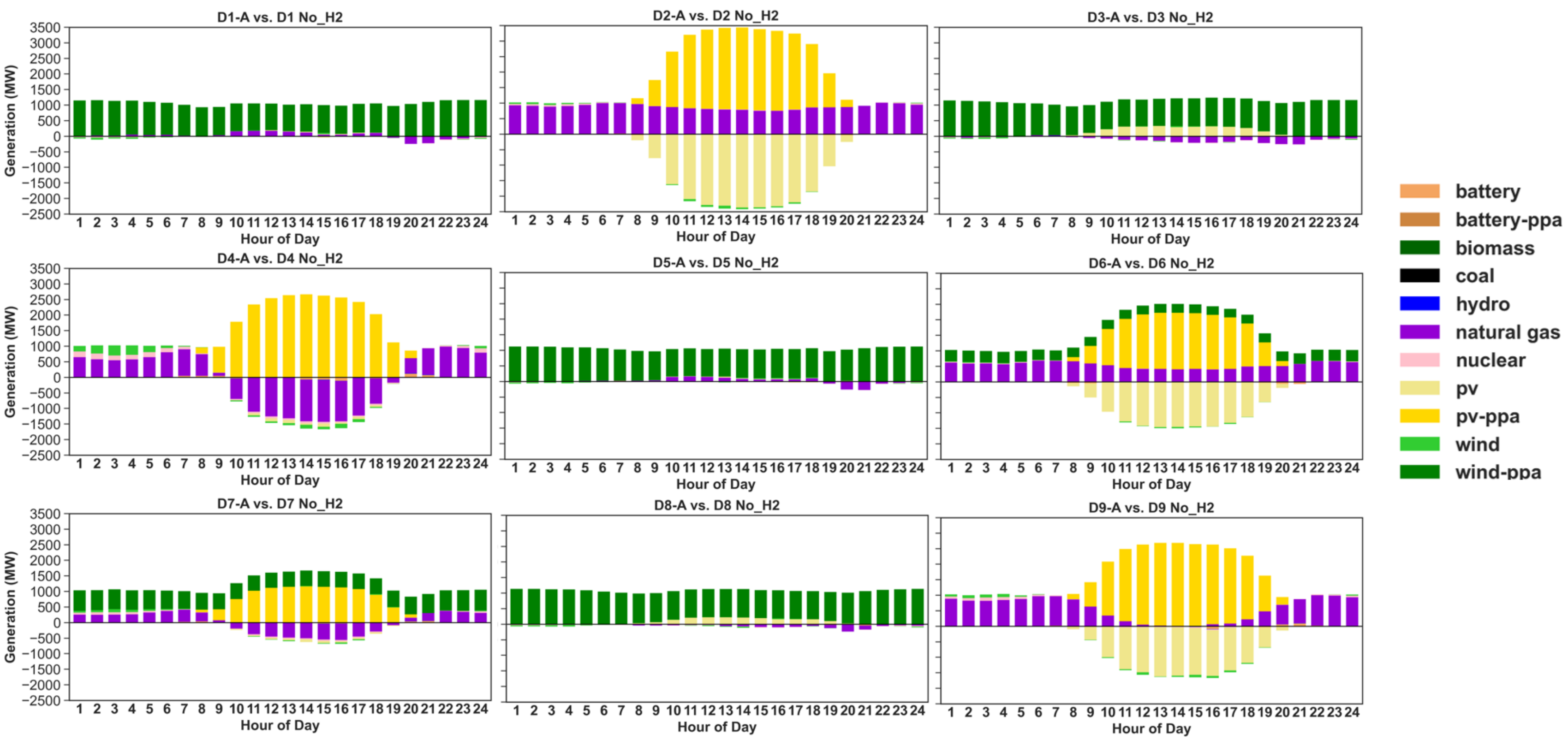

*Figure S 1. Difference in average hourly dispatch between counterfactual (with $H_2$ demand and time matching requirement –"DX-A") and baseline grid without $H_2$ demand for the deterministic model under different weather scenarios and annual time matching. Each panel is labeled to show results for one weather scenario. DX-A = Deterministic model using weather scenario "X" (1-9) with $H_2$ production and annual time matching. DX No_H2 = Deterministic using weather scenario "X" (1-9) without $H_2$ production and time matching requirement.*

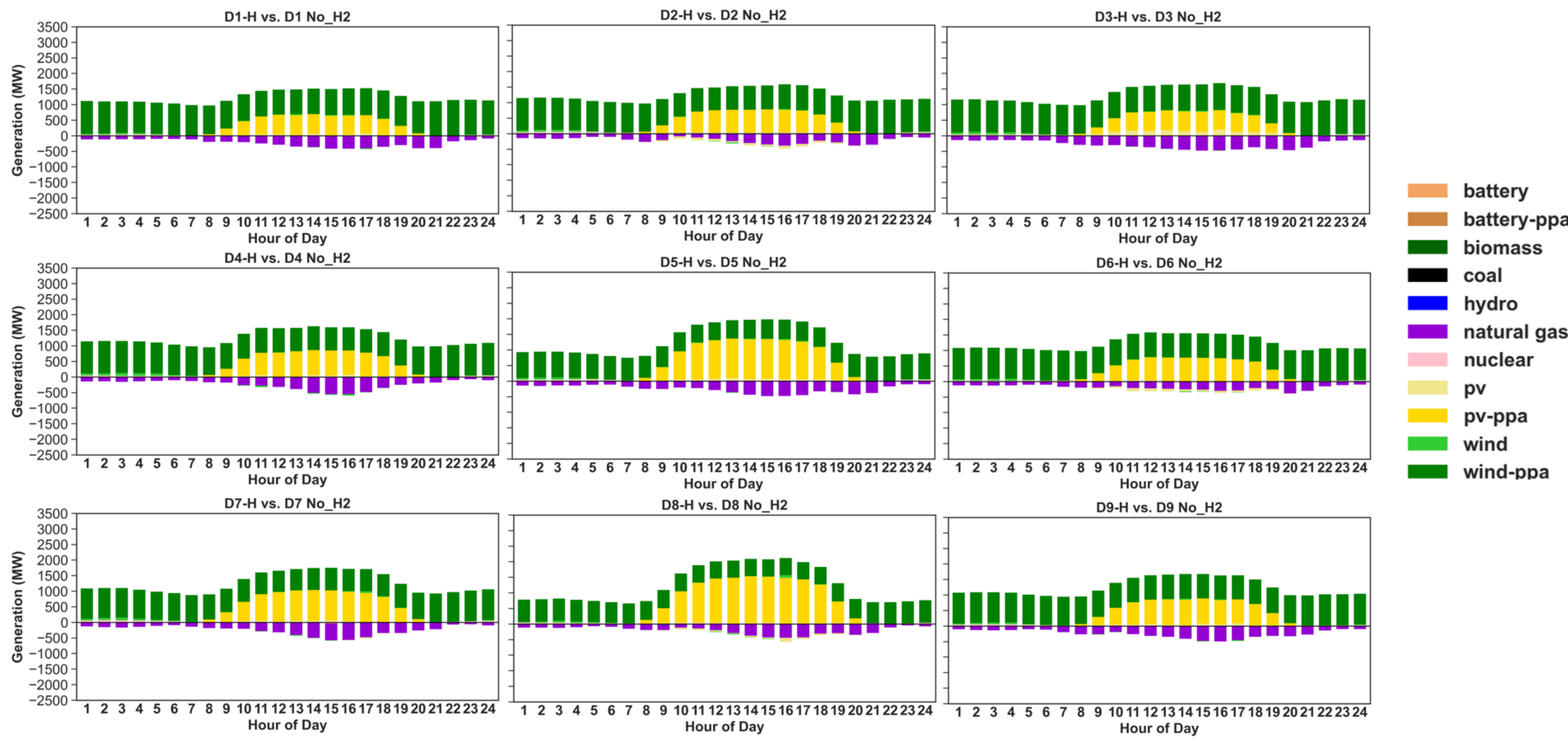

*Figure S 2. Difference in average hourly dispatch between counterfactual (with $H_2$ demand and time matching requirement –"DX-A") and baseline grid with no $H_2$ demand for the deterministic model under different weather scenarios and hourly time matching. Each panel is labeled to show results for one weather scenario. DX-H = Deterministic model using weather scenario "X" (1-9) with $H_2$ production and annual time matching. DX No_$H_2$ = Deterministic using weather scenario "X" (1-9) without $H_2$ production and time matching requirement.*

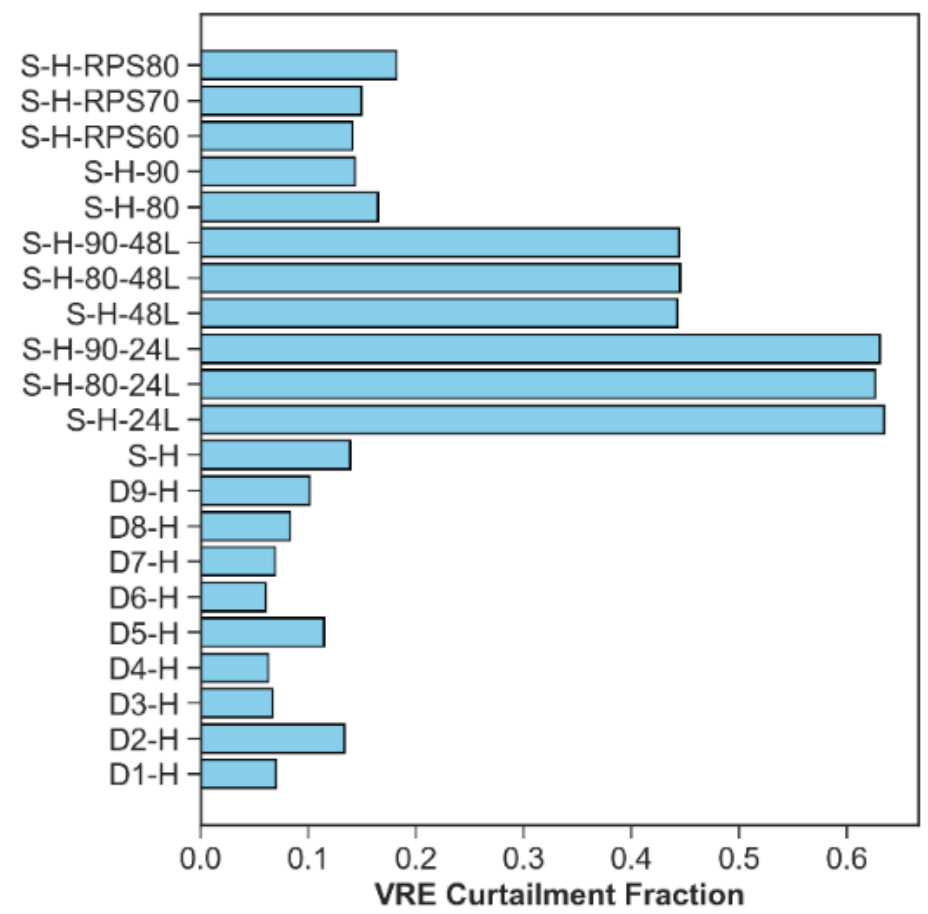

*Figure S 3. VRE curtailment fraction for PPA resources across deterministic and stochastic model cases with hourly matching. VRE curtailment defined as the ratio of curtailed energy production from VRE resources to the available energy from VRE resources. DX-H= Deterministic model for weather scenario X (1-9) with hourly matching; S-H= Stochastic model with hourly matching; S-H-X = Stochastic model with hourly matching with X percent (80,90) compliance of the constraint; S-H-X-YL = Stochastic model with hourly matching with X percent compliance (80,90) and limit on $H_2$ storage capacity equal to Y hours of $H_2$ demand (Y = 24, 48). S-H-YL= Stochastic model with hourly matching with 100 percent compliance and limit on $H_2$ storage capacity equal to Y hours of H2 demand (Y = 24, 48). Annual matching scenarios generally have near zero VRE curtailment and hence are not shown.*

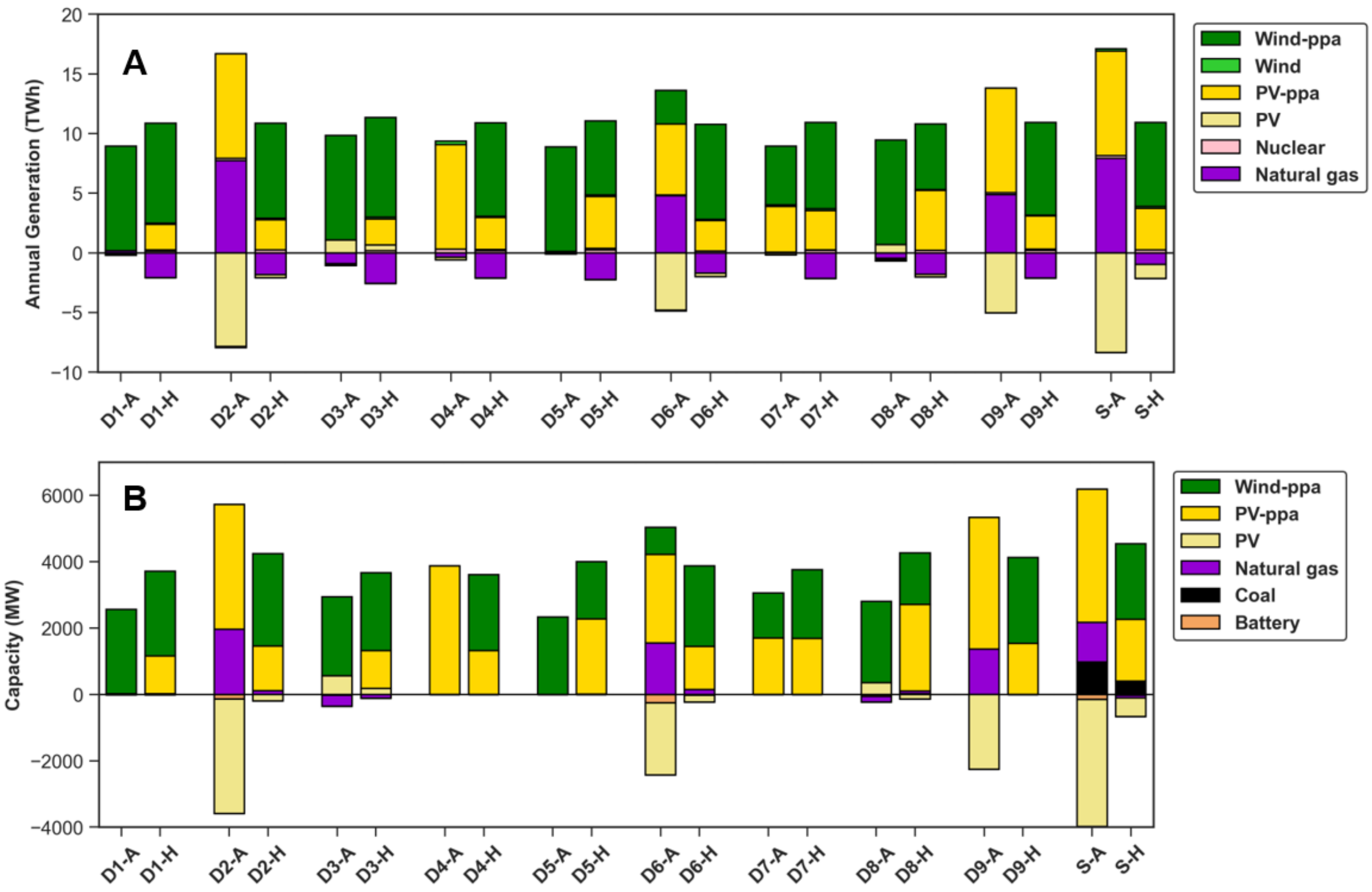

*Figure S 4. Difference in annual generation (A) and power generation and storage capacity (B) as a result of electricity-based $H_2$ production with annual or hourly TMR under deterministic and stochastic models. Model configuration nomenclature as defined in Figure 1 caption.*

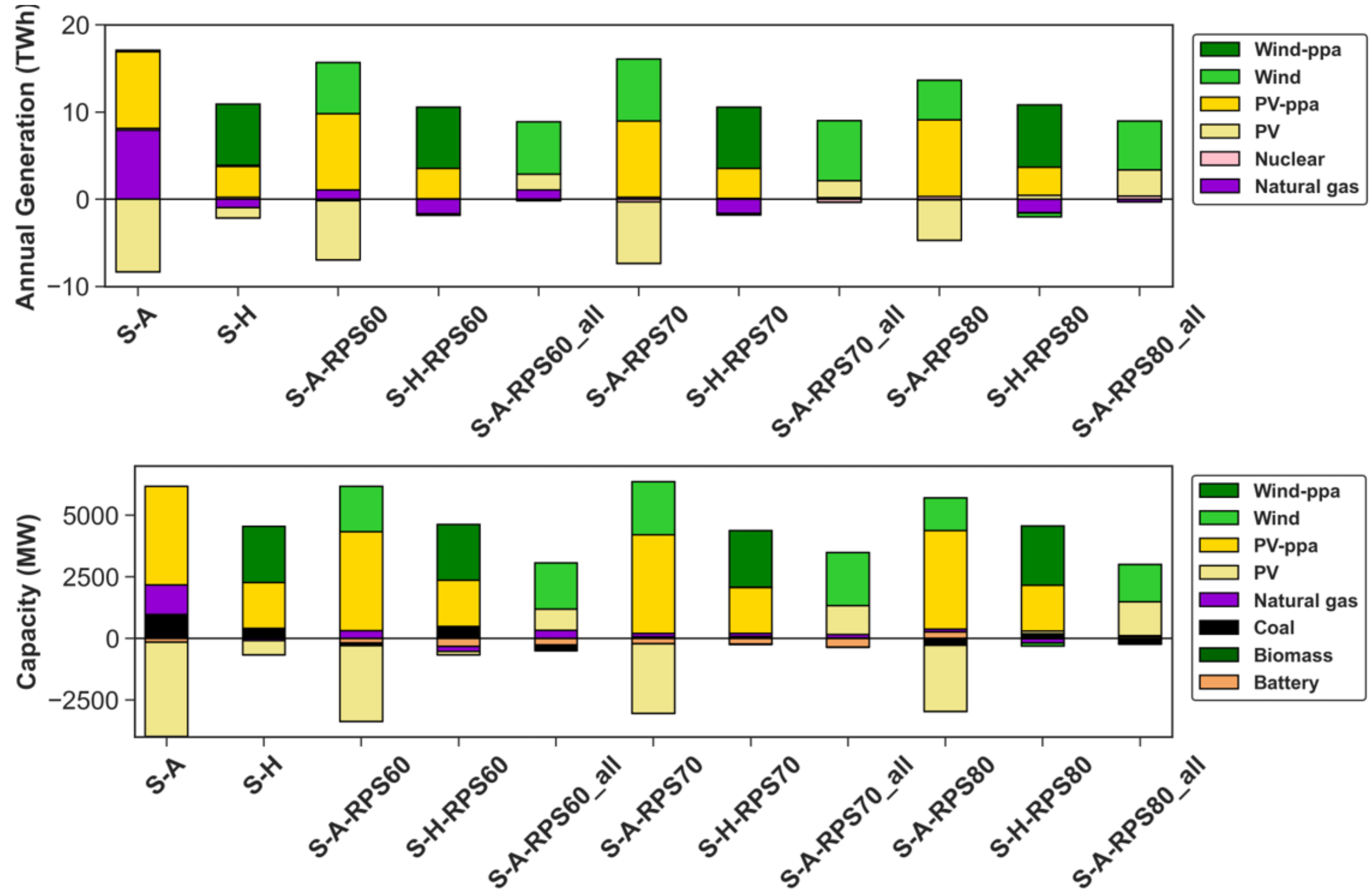

*Figure S 5. Difference in annual generation mix (top) and capacity mix (bottom) with and without electricity demand from $H_2$ production under different scenarios of time-matching requirements on H2-related electricity matching (no, annual, hourly) and renewable portfolio standard (RPS) requirements (no RPS, 60%, 70%, 80%) on non-$H_2$ electricity demands. "RPS 60" indicates 60% minimum annual VRE generation for non-$H_2$ demand and so on. "RPSX_all" includes electrolyzer consumption in the RPS constraint with no time-matching requirement. S-A/S-H: annual/hourly time-matching and no RPS constraint. S-A-RPSXX/S-H-RPSXX: annual/hourly time-matching for electrolyzer with XX% RPS for non-$H_2$ demand.*

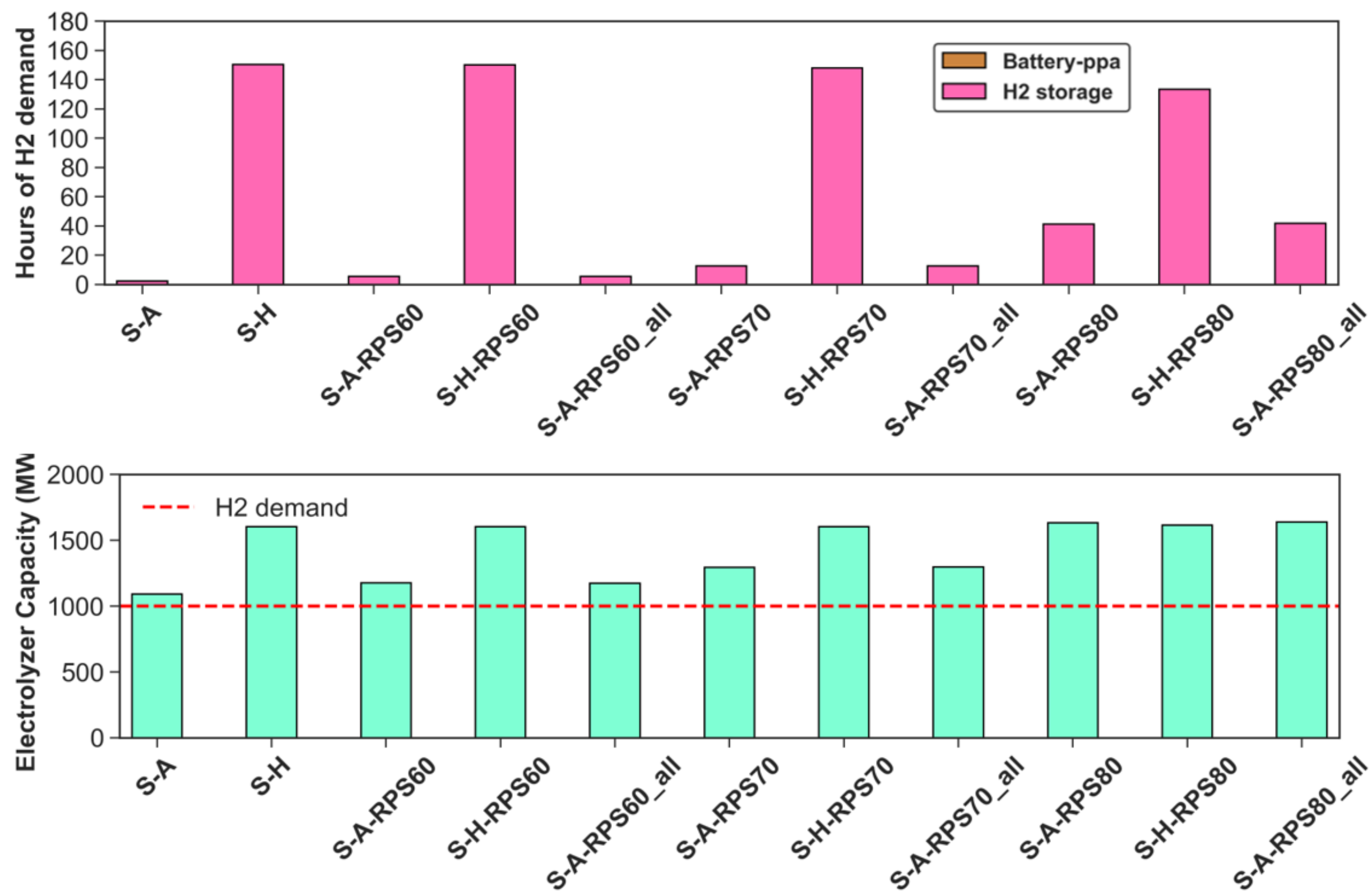

*Figure S 6. Installed PPA energy storage capacity (top) and electrolyzer capacity (bottom) to support electrolytic $H_2$ production (1 GW) under different scenarios of time-matching requirements (no, annual, hourly) and renewable portfolio standard (RPS) requirements (no RPS, 60%, 70%, 80%) on non-$H_2$ electricity demands. "RPS 60" indicates 60% minimum annual VRE generation for non-$H_2$ demand and so on. "RPSX_all" includes electrolyzer consumption in the RPS constraint with no time-matching requirement. S-A/S-H: annual/hourly time-matching and no RPS constraint. S-A-RPSXX/S-H-RPSXX: annual/hourly time-matching for electrolyzer with XX% RPS for non-$H_2$ demand. The $H_2$ storage capacity is reported in terms of hours of $H_2$ demand, which is calculated by dividing the $H_2$ storage capacity by the baseload $H_2$ demand (18.4 tonnes/hour). No PPA battery energy storage is deployed across the evaluated scenarios. All results shown are for the stochastic model.*

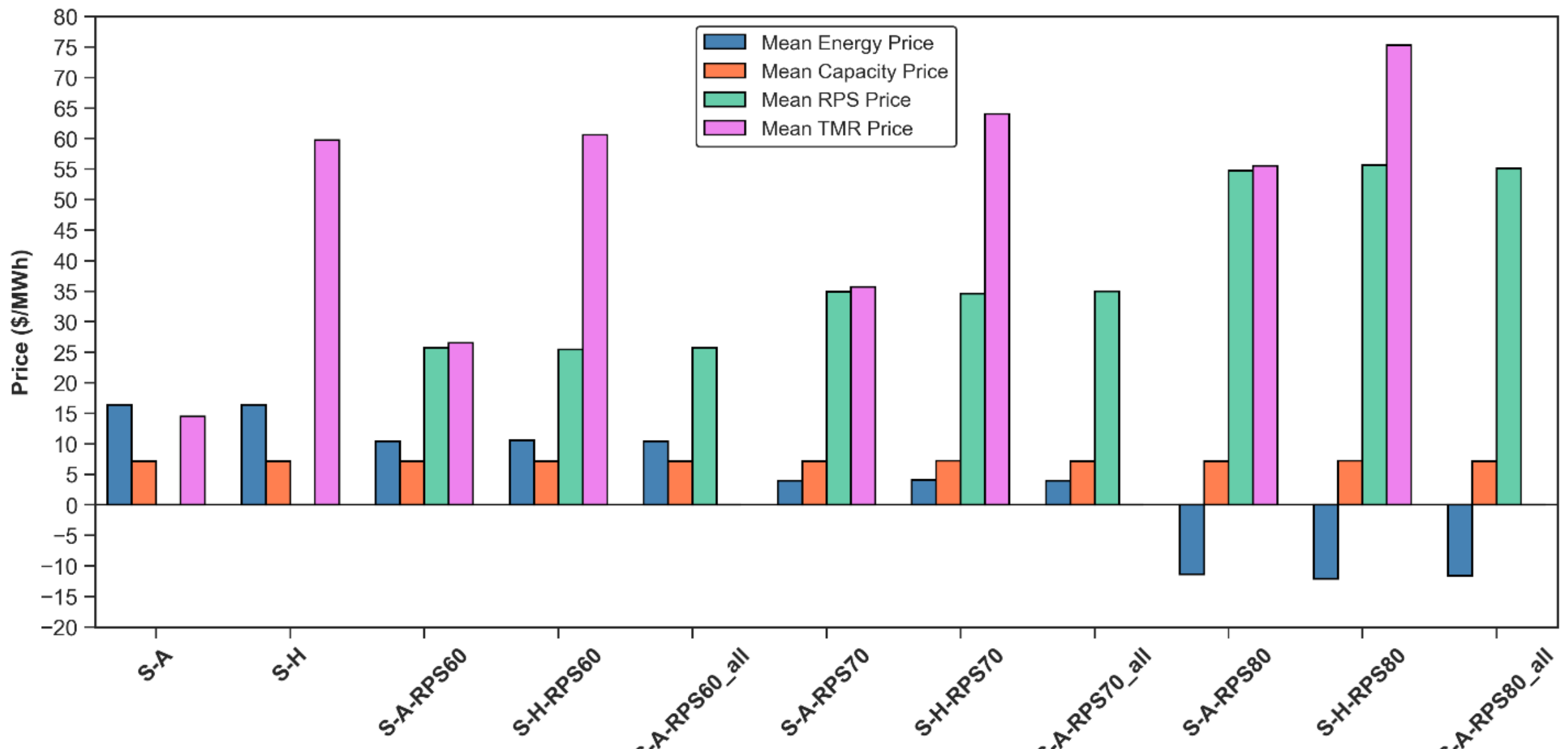

*Figure S 7. Average energy price, capacity price, renewable portfolio standard (RPS) price, and time-matching requirement (TMR) price for stochastic model runs with different time-matching requirements (TMR) (no TMR, annual, hourly) as well as RPS constraints (no RPS, 60%, 70%, 80%). See Figure S 5 caption for scenario name explanations. Energy price – shadow price of hourly electricity supply-demand balance constraint, capacity price – shadow price of hourly resource adequacy constraint, RPS price – shadow price of annual RPS constraint, TMR price – shadow price of TMR constraint (annual or hourly depending on the case). RPS = Renewable Portfolio Standard.*

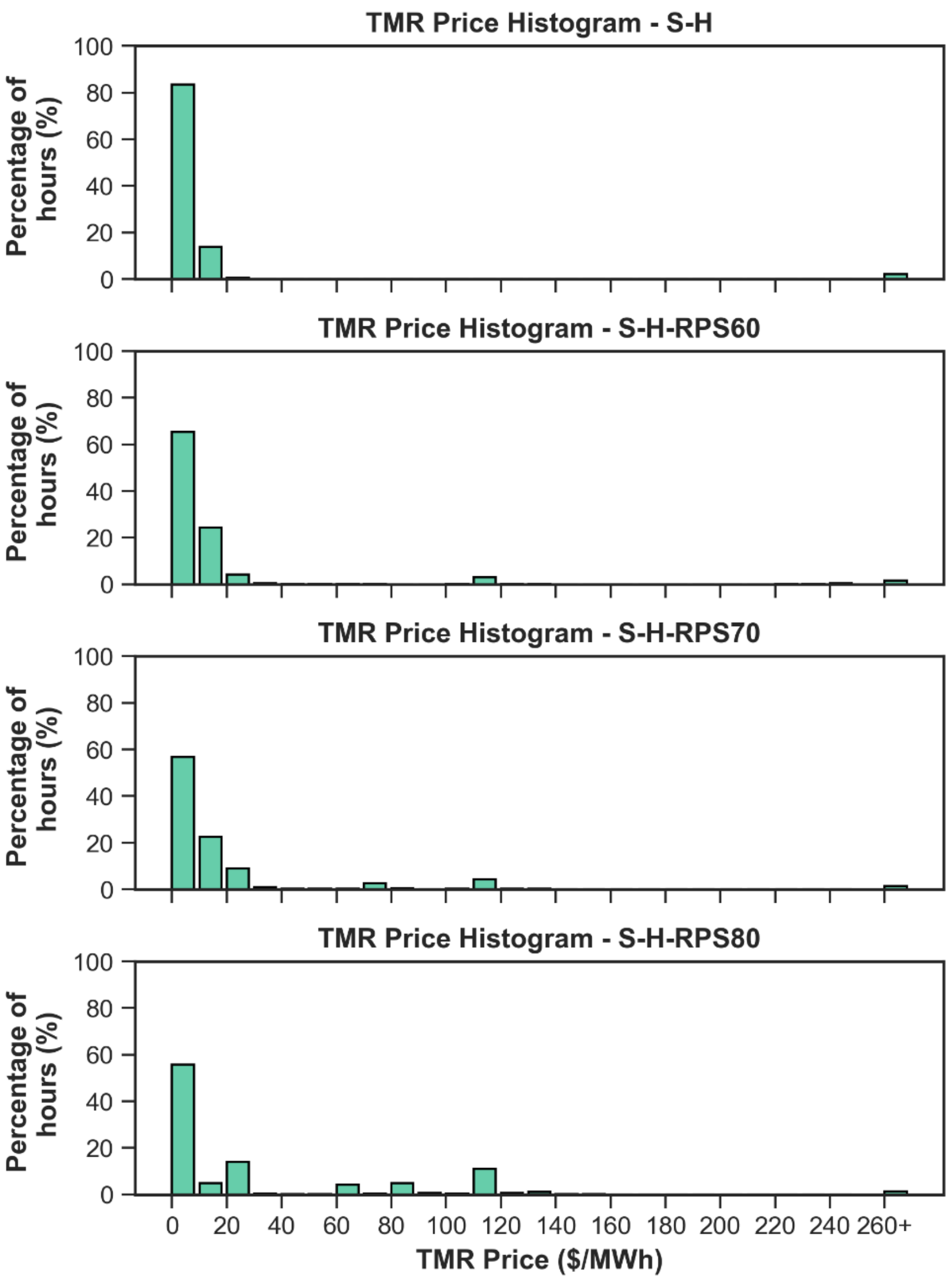

*Figure S 8. Distribution of hourly time-matching requirement constraint shadow prices for the stochastic model evaluated under different scenarios of RPS requirement for non-$H_2$ (e.g., grid) electricity demand. Prices greater than $260 /MWh are aggregated in the last bin labeled "260+".*

# S3. Model data inputs

## S3.1. VRE Data and Scenario Selection

Both the deterministic and stochastic model use hourly VRE availability profiles from ERCOT as inputs. The deterministic model considers one year of hourly VRE generation (8760 hours), whereas the stochastic model considers nine years of hourly data (9 * 8760 hours) for each technology (wind, solar) by resource type (existing and new).

To construct these profiles, we use the ERCOT's Hourly Wind and Solar Generation Profiles dataset[4] which provides hourly solar and wind generation profiles for existing and planned plants from 1980 to 2021. Existing plants are defined as VRE plants that were operational as of 2020. Planned plants are VRE plants that had received approval for or were under construction as of 2020. The ERCOT dataset uses spatially granular historical weather data to estimate hourly generation from both types of resources for all years in the dataset — e.g., a planned resource will still have an hourly generation profile available for 1980 that is based on the technical parameters of that plant and the weather conditions in 1980. To construct the VRE profiles input into our single-region model of ERCOT, we aggregate by existing or planned resources, sum the hourly generation within both groups, and divide the aggregated hourly generation by the total capacity of each group. The result is four time series (existing/planned x wind/solar) of hourly capacity factors for the years 1980-2021(Figure S 9**Error! Reference source not found.**).

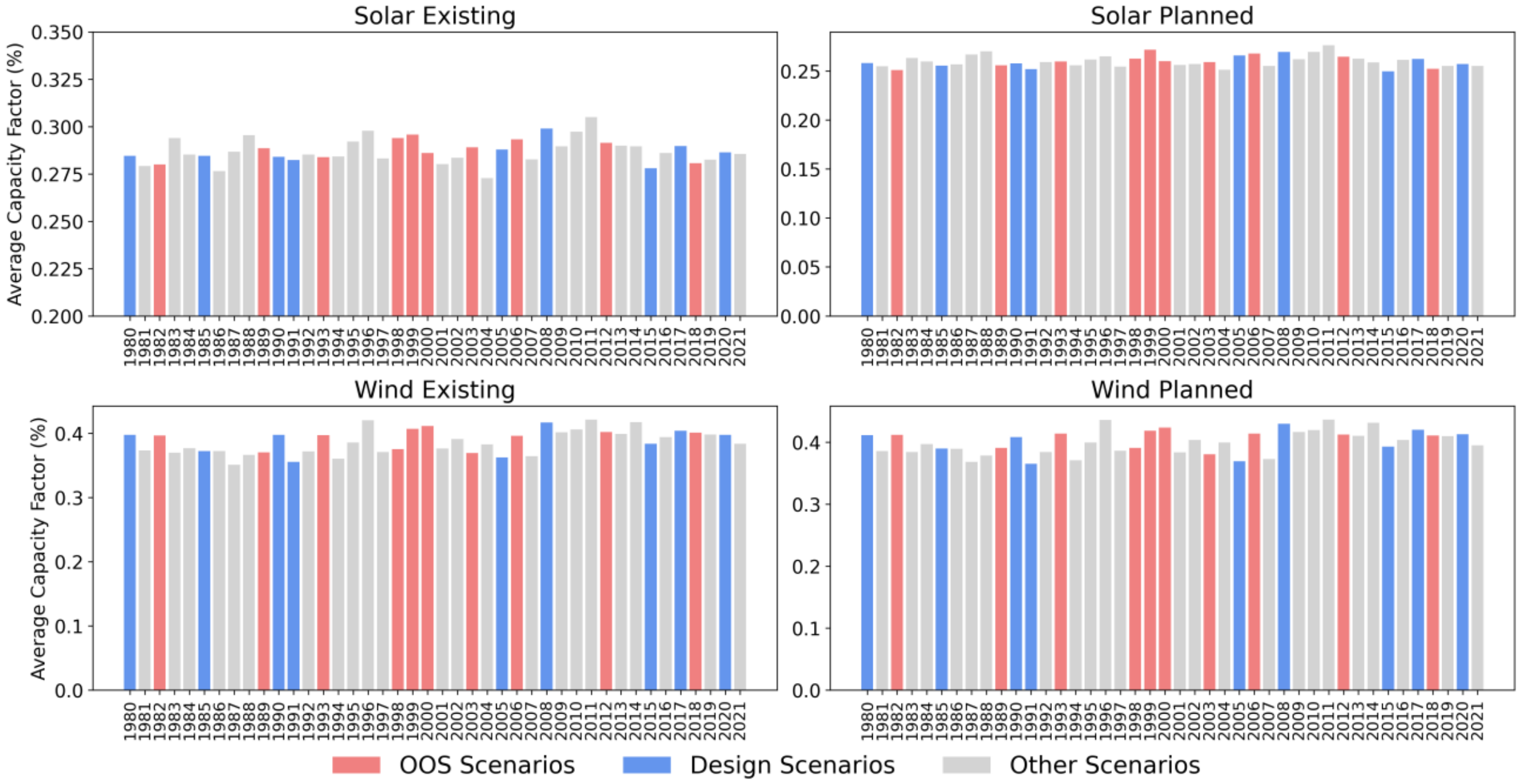

*Figure S 9.* ***Annual averaged capacity factors of solar and wind resources in ERCOT (1980-2021).*** *Capacity factors are reported for solar (A, B) and wind (C, D) resources that are either existing (A, C) or planned (B, C) as of 2021. Note that the y-axes do not extend to zero, which is done to make it easier to observe variation among years. OoS = Out of sample; Only OoS and Design scenarios are used in the analysis.*

To balance the computational resource needs to solve the stochastic model while striving to maximize the extent to which we capture the variability among VRE availability profiles, we perform scenario reduction via k-means clustering to identify nine representative scenarios from the data. As seen in other studies[1,2,5], wind is favored to meet an hourly time-matching requirement, so we select representative VRE years based on wind. Solar scenarios correspond to the years selected from the k-means clustering for wind. The nine representative VRE scenarios correspond to the years 1980, 1985, 1990, 1991, 2005, 2008, 2015, 2017, and 2020, whose hourly capacity factor distribution for new wind and solar resources is highlighted in Figure S 10A and B, respectively.

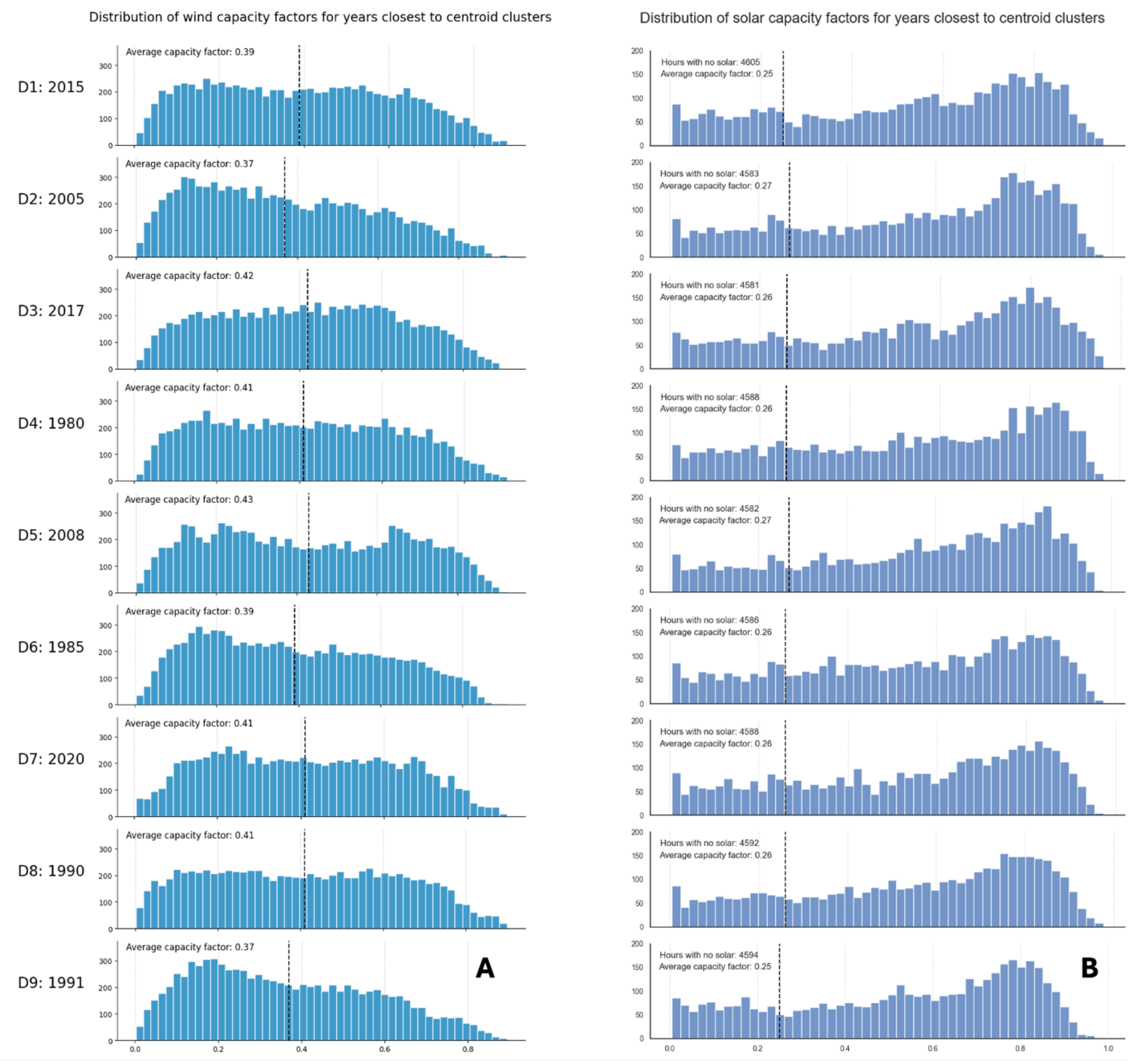

*Figure S 10.* ***Distribution of hourly wind and solar capacity factors for the nine design cases.*** *Data corresponds to planned wind resources (A) and planned solar resource (B) from the nine VRE years selected via k-means clustering from the 41 ERCOT VRE scenarios. Design scenarios are label "DX" followed by the year of ERCOT's VRE data that they correspond to. Vertical lines indicate the average capacity factor. To make it easier to see the distribution of hours with non-zero capacity factors for solar, hours with capacity factors of less than 0.005 are not shown in the chart, but the number of such hours is reported as "Hours with no solar" in the top left of each subplot.*

The 10 out-of-sample VRE scenarios were selected by randomly sampling from the 31 VRE scenarios that are not used for the design cases. The selected years were 1982, 1989, 1993, 1998, 1999, 2000, 2003, 2006, 2012, and 2018 whose hourly capacity factor distribution for new wind and solar resources is highlighted in Figure S 11A and B, respectively.

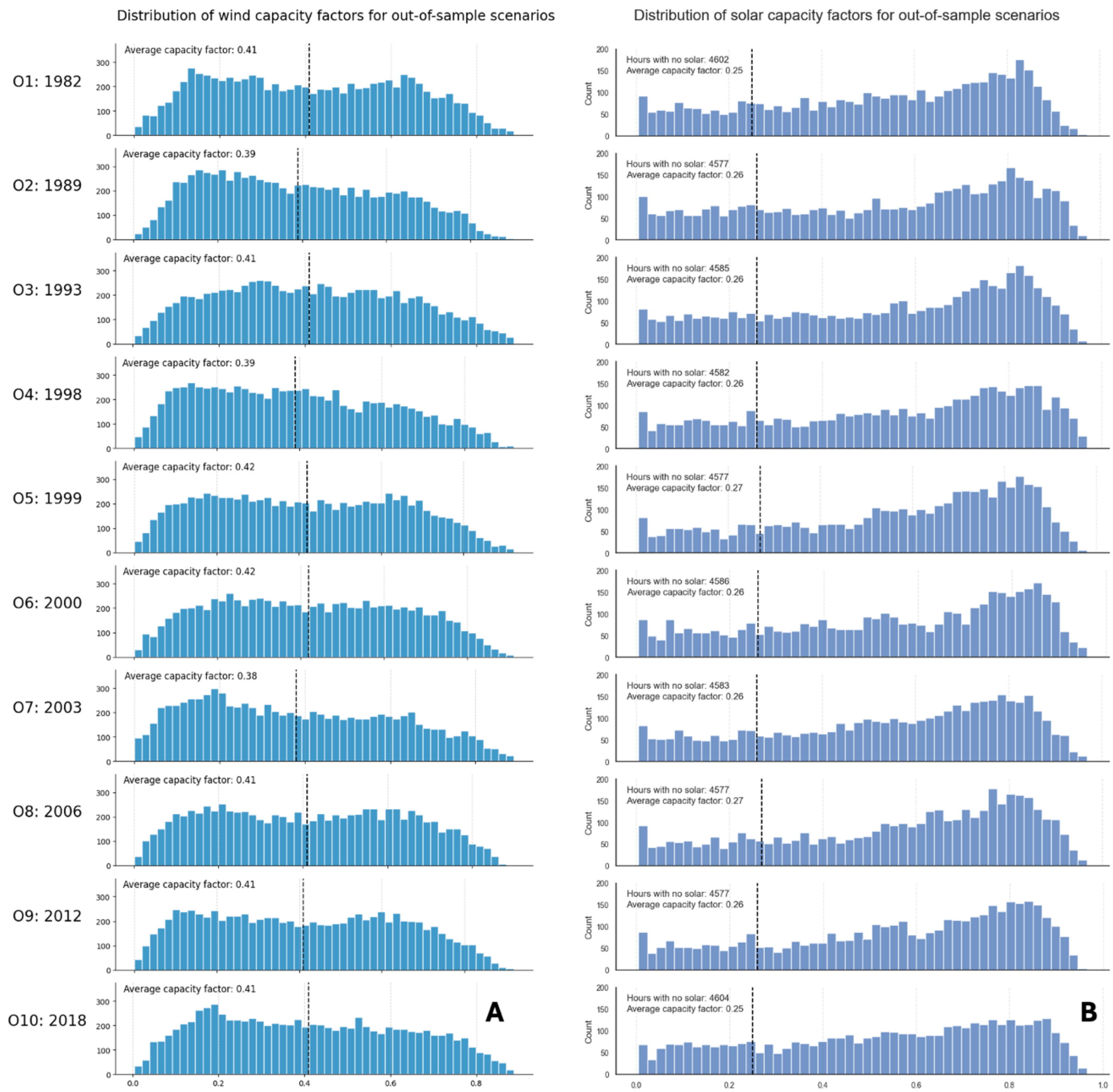

*Figure S 11.* ***Distribution of hourly wind and solar capacity factors for the 10 out-of-sample scenarios.*** *Data corresponds to planned wind resources (A) and planned solar resource (B) from the 10 VRE years randomly selected from 31 ERCOT VRE scenarios, excluding the years used for the design cases. Out-of-sample scenarios are label "OX" followed by the year of ERCOT's VRE data that they correspond to. See caption of Figure S9 for further details.*

## S3.2. Technology cost and performance assumptions

This section summarizes the major technology cost and performance assumptions. Unless otherwise specified, the stochastic model and deterministic model use the same assumptions as described in our prior work[6]— i.e., technology cost and performance, as well as power system characteristics, such as electricity demand, existing generators, value of lost load.

All costs are reported on a 2021 USD. Table S2 summarizes the cost assumptions for VRE, natural gas and Li-ion battery storage resources. Battery storage has a self-discharge rate of 0.002% per hour[7]. The model can independently vary the installed energy capacity and power capacity for Li-ion storage so long as the ratio of energy capacity to power capacity (i.e., duration) is between 0.15-12 hours. Table S3 summarizes cost assumptions for electrolyzers and $H_2$ storage. Fuel prices correspond to 2019 data adjusted to 2021 USD, due to COVID-related disruptions to fuel prices that occurred in 2021.

*Table S2. Electricity generation and storage technology cost and operational parameters. Annualized investment costs are calculated with a discount rate of 4%. Cost assumptions are adapted from the NREL Annual Technology Baseline (ATB) 2022 edition[8]. The capex values reported in the table include a 30% discount to approximate the investment tax credit for wind, solar and storage consistent with policy as per the U.S. inflation reduction act of 2022. Li-ion battery storage is subject to a variable operation and storage cost to discourage simultaneous charging and discharging, which must be explicitly discouraged a linear programming capacity expansion model. NGCT = Natural Gas combustion turbine. NGCC = Natural gas combined cycle.*

| Technology | | Solar PV | Onshore wind | Li-ion battery storage | NGCT | NGCC |
|---|---|---|---|---|---|---|
| Lifetime (years) | | 30 | 30 | 15 | 30 | 30 |
| Investment cost – power ($/MW) | | 1,176,000 | 1,428,000 | 255,150 | 950,249 | 1,080,449 |
| Annualized CAPEX – Power ($/MW/year) | | 47,606 | 57,807 | 16,064 | | |
| Investment cost – energy ($/MWh) | | - | - | 296,100 | | |
| Annualized CAPEX – Energy ($/MWh/year) | | - | - | 18,642 | - | |
| Fixed operation and maintenance cost | Power ($/MW/year) | 21,000 | 44,100 | 6,379 | 11,849 | 13,513 |
| | Energy ($/MWh/year) | - | - | 7,403 | | |
| Variable operating cost ($/MWh) | | 0.00 | 0.00 | 1.00 | 5.00 | 2.00 |
| Heat rate (MMBtu/MWh) | | - | - | - | 9.71 | 6.36 |
| Start-up fuel use (MMBtu per start) | | - | - | - | 815.5 | 1454.0 |
| Start cost ($/start) | | - | - | - | 27,028 | 64,703 |

*Table S3. Hydrogen electrolyzer and storage cost and operation parameters. Annualized investment costs are calculated with a discount rate of 4%. Electrolyzer assumptions reflect 2022 assumptions from the NREL H2A report[9] , and hydrogen storage assumptions are from Papadias and Ahluwalia (2021). Costs reported in $/MWh are calculated by multiplying $/tonne H2 by the lower heating value of hydrogen. Water costs are assumed to be negligible and therefore are not considered. t = tonne; y = year.*

| **Technology** | **Lifetime** | **Investment cost** | | **Annualized investment cost** | | **Fixed operation and maintenance (FOM) cost -$H_2$ production rate** $(\$/\mathrm{MWH_2}/\mathrm{year})$ | **Electrical power use** $(\mathrm{MWh/t\ H_2})$ |
|---|---|---|---|---|---|---|---|
| | | **H2 production rate** $(\$/\mathrm{MWH_2})$ | **Energy** $(\$/\mathrm{t\ H_2})$ | **$H_2$ Production rate** $(\$/\mathrm{MWH_2/y})$ | **Energy** $(\$/\mathrm{t\ H_2/y})$ | | |
| Electrolyzer | 20 | 1,937,791 | - | 142,586 | - | 28,604 | 54.3 |
| $H_2$ storage (tank) | 30 | - | 587,000 | - | 33,929 | - | - |
| $H_2$ storage compressor | 15 | 2,451,496 | - | 220,490 | - | - | 0.71 |

*Table S4. Global parameters. Fuel prices are from the EIA 2022 Annual Energy Outlook 2022. Natural gas and coal modeled with $CO_2$ emissions factors of 0.05306 $tCO_2$/MMBtu and 0.09552 $tCO_2$/MMBtu, respectively. The capacity reserve margin is selected based on minimum target used by ERCOT for planning purposes. Slack penalties are reported for all constraints where a slack variable is applied. The high penalty for unmet $H_2$ and power demand was selected to avoid any load shedding instances. The unserved grid load penalty is based on the value of lost load previously used in ERCOT.*

| | **Value** | **Applicable Cases** |
|---|---|---|
| **Discount Rate used for annualization of capital costs** | 4% | All |
| **Capacity Reserve Margin (used in Eq. S1)** | 13.75% | All |
| **Fuel Prices ($/MMBtu)** | | |
| Natural Gas | 2.03 | All |
| Coal | 2.47 | All |
| Uranium (nuclear) | 0.70 | All |
| **Penalties for slack variables in various constraints** | | |
| Unserved $H_2$ demand ($/tonne $H_2$) | 5e+07 | All |
| Unserved power demand ($/MWh) | 9,000 | All |
| Unmet RPS requirement ($/MWh) | 1,000 | RPS cases |
| Unmet Time-matching requirement ($/MWh) | 500 | Out-of-sample cases with a hourly time-matching requirement for hydrogen |

*Table S5. Capacity reserve margin derating factors by resource type. Source: prior modeling study[10]. Refer to Eq. S1 and Github repository[3] for implementation of constraint.*

| | Capacity Reserve Derate Factor (see Eq. S1) |
|---|---|
| Coal | 0.93 |
| Natural gas combined cycle | 0.93 |
| Natural gas combustion turbine | 0.93 |
| Nuclear | 0.93 |
| NG steam turbine | 0.93 |
| Biomass | 0.90 |
| Hydro | 0.80 |
| Solar | 0.80 |
| Wind (onshore) | 0.80 |
| Diurnal battery storage | 0.80 |

## S3.3 Existing generator fleet

To increase the number of weather scenarios considered in the stochastic model while maintaining computational tractability with off-the-shelf LP solvers (e.g., Gurobi), we reduced the resolution of the characterization of the existing power generation fleet, sourced from the PowerGenome data base[11]. Specifically, we combined all coal and natural gas steam turbines that either operated at <5% capacity factor in the baseline run (i.e. without any $H_2$ demand) from the previous analysis or had heat rates greater than 15 MMBTU/MWh into one cluster. This reduced the total number of generators (including existing and candidate new) from 64 to 49, which enabled more weather scenarios to be considered in the stochastic model.

*Table S6. Capacity and maximum hourly availability of generators. Dispatchable fossil, nuclear, and biomass generators are assumed to experience outages, maintenance, etc. resulting in less than 100% availability at all hours. Hydro, solar, and wind availability are assumed to be subject to weather conditions and therefore their maximum hourly variability is time-dependent and separately specified.*

| | Capacity (GW) | Maximum Hourly Availability Factor (%) |
|---|---|---|
| Coal | 14.4 | 90 |
| Natural gas combined cycle | 35.13 | 90 |
| Natural gas combustion turbine | 6.83 | 90 |
| Nuclear | 4.98 | 95 |
| Natural gas steam turbine | 10.23 | 90 |
| Biomass | 0.07 | 90 |
| Hydro | 0.5 | N/A |
| Solar | 9.14 | N/A |
| Wind (onshore) | 34.06 | N/A |